\documentclass[12pt,a4paper]{article}            
 \usepackage[skins,theorems]{tcolorbox}
\tcbset{highlight math style={enhanced,
  colframe=red,colback=white,arc=0pt,boxrule=1pt}}
  \usepackage[bookmarksopen, bookmarksnumbered, bookmarksopenlevel=2]{hyperref}
  \usepackage{tikz}
  \usepackage{tikz-3dplot}
  \usepackage{multirow}
 \usetikzlibrary{calc}
 \usetikzlibrary{decorations} %
 \usepackage[UKenglish]{babel}
 \usepackage[toc,page]{appendix}
 \usepackage{amsmath}
 \usepackage{amssymb}
 \usepackage{graphicx}
 \usepackage{hhline}
 \usepackage[bf]{caption}
\usepackage{cite}
\usepackage[vcentermath]{youngtab}
\usepackage{geometry}
\usepackage{slashed}
\usepackage{color}
\usepackage{stackrel}
\usepackage{tikz-cd} 
\usepackage{cancel} 
\usepackage[normalem]{ulem}
\usepackage{empheq}
\usepackage{arydshln}
\usepackage{tablefootnote}

\newenvironment{eqn*}{\begin{equation*}\begin{aligned}}{\end{aligned}\end{equation*}\noindent}
 \topmargin
-1.5cm
\textwidth
15.5cm
\textheight
23.5cm
\oddsidemargin
0.7cm
\evensidemargin
0.7cm

\newcommand{\bqa}{\begin{eqnarray}}
\newcommand{\eqa}{\end{eqnarray}}

\hypersetup{
    pdftitle={},
    pdfauthor={},
    pdfsubject={}
}
\numberwithin{equation}{section}
\numberwithin{table}{section}

\makeatletter

\newcommand{\be}{\begin{equation}}
\newcommand{\ee}{\end{equation}}
\newcommand{\beq}{\begin{equation}}
\newcommand{\eeq}{\end{equation}}
\newcommand{\ba}{\begin{aligned}}
\newcommand{\ea}{\end{aligned}}

\newcommand{\bea}{\begin{eqnarray}}
\newcommand{\eea}{\end{eqnarray}}

\newcommand{\cO}{\mathcal{O}}
\newcommand{\cT}{\mathcal{T}}
\newcommand{\cE}{\mathcal{E}}

\newcommand{\cC}{\mathcal{C}}

\newcommand{\cN}{\mathcal{N}}

\newcommand{\cA}{\mathcal{A}}

\newcommand{\cM}{\mathcal M}

\newcommand\bi{\begin{itemize}}
\newcommand\ei{\end{itemize}}

\def\Im{\mathop{\mathrm{Im}}\nolimits}

\def\Tr{\mathop{\mathrm{Tr}}\nolimits}
\def\tr{\mathop{\mathrm{tr}}\nolimits}

\def\unit{{1\kern-.65ex {\rm l}}}
\def\1{{1\kern-.65ex {\rm l}}}

\def\CE{{\cal E}}

\def\CZ{{\cal Z}}

\newcount\hour \newcount\minute
\hour=\time \divide \hour by 60
\minute=\time
\def\now{%
\ifnum \hour<13
  \ifnum \hour=0 \advance \hour by 12 \number\hour:\else \number\hour:\fi%
     \ifnum \minute<10 0\fi%
     \number\minute%
\ A.M.%
\else \advance \hour by -12 \number\hour:%
  \ifnum \minute<10 0\fi%
  \number\minute%
  \ P.M.%
\fi%
}

\makeatother

\begin{document}

\vskip 40 pt
\begin{center}

\rightline{ \small IFT-UAM/CSIC-25-117}
\vspace{1.4cm}

{\large \bf
 Towards the Non-Perturbative Completion of 4d $\mathcal N=1$ \vspace{0.5em}\newline
 Effective Theories of Gravity
} 

 \vskip 11 mm

Gonzalo F. Casas,${}^1$  Max Wiesner${}^{2}$

\vskip 11 mm
${}^1${\it Instituto de Fisica Teorica UAM-CSIC, c/ Nicolas Cabrera 13-15, 28049 Madrid, Spain}  \\[3 mm]

${}^2$\textit{II. Institut f\"ur Theoretische Physik, Universit\"at Hamburg, Notkestrasse 9,\\ 22607 Hamburg, Germany}

\end{center}

\vskip 7mm

\begin{abstract}
We show that four-dimensional $\mathcal N=1$ effective theories of gravity obtained from string compactifications require a non-perturbative completion, as additional light states of non-perturbative origin must be incorporated in the small volume regime to obtain a consistent low-energy description. 
This completion becomes concrete in subsectors that locally exhibit enhanced supersymmetry, where the enhancement predicts the existence of additional light degrees of freedom absent in the perturbative description. Motivated by analogous setups in six-dimensional $\mathcal N=(1,0)$ theories, we focus on the Kähler moduli space of F-theory compactifications on Calabi--Yau fourfolds to four dimensions, where shrinkable curves not intersected by 7-planes realize such supersymmetry-enhanced subsectors. Guided by the enhanced supersymmetry, we use F-theory to identify the degrees of freedom missing in the small volume regime of the perturbative Type IIB description. A consistent embedding of these local subsectors into a four-dimensional $\mathcal N=1$ theory of gravity requires an appropriate inclusion of complex structure moduli and spacetime-filling D3-branes.  We also discuss supersymmetry enhancement in the complex structure sector and study how a heterotic dual description gives a unifying picture of the different F-theory sectors with enhanced supersymmetry. Finally, we comment on cases without local supersymmetry enhancement.

\end{abstract}

\vfill

\thispagestyle{empty}
\setcounter{page}{0}
 \newpage

\thispagestyle{empty}
\tableofcontents
\vspace{25pt} 
\setcounter{page}{1}

\section{Introduction \& Summary}
A major open question in the study of theories of quantum gravity concerns the role of non-perturbative effects for the consistency of such theories.  This is particularly relevant for identifying general consistency conditions valid for any theory of quantum gravity as they have been formulated in the context of the Swampland program~\cite{Vafa:2005ui}, see \cite{Palti:2019pca,Grana:2021zvf,vanBeest:2021lhn,Agmon:2022thq} for reviews. While the framework of perturbative string theory provides a rich arena for probing quantum gravity, it is also known that non-perturbative effects are crucial for consistently connecting theories realized in different corners of the quantum gravity landscape. 

A classic example illustrating this point can be found in compactifications of Type II string theory on Calabi--Yau threefolds, which lead to an effective 4d $\cN=2$ supergravity theory in the low-energy limit. The massless degrees of freedom of these theories assemble into vector and hypermultiplets, with the number of such multiplets being determined by the topology of the chosen threefold. For 4d $\cN=2$ theories, supersymmetry ensures that the moduli space $\cM_{\cN=2}$ of the effective theory factorizes into the vector multiplet moduli space $\cM_{\rm VM}$ and the hypermultiplet moduli space $\cM_{\rm HM}$. Different Calabi--Yau threefolds can be connected by extremal transitions that change their topology. Examples of such transitions include conifold transitions~\cite{Green:1988bp,Candelas:1988di,Green:1988wa,Candelas:1989ug,Aspinwall:1993nu,Aspinwall:1993yb}, which result in changes to the number of vectors and hypermultiplets in the resulting low-energy effective action. Geometrically, these transitions are possible at loci in the deformation space of the threefold along which certain cycles shrink to zero size while keeping the overall size of the threefold finite. The physics associated with the shrinking curve can then be described by a field theory sector that decouples from gravity in the vicinity of the transition point. From a field theory perspective, the geometric transition occurs when moving from the Coulomb branch to the Higgs branch, or vice versa, with the geometrically singular locus corresponding to the origin of either branch.  As the origin of the field theory branches correspond to cycles vanishing in the Calabi--Yau threefold, this regime of the field space cannot be fully described within the supergravity approximation of string theory. In other words, perturbative string theory does not fully account for the physics associated with geometric transitions~\cite{Strominger:1995cz,Greene:1995hu,Ooguri:1996me}. 

In Type II compactifications, there are two ways that non-perturbative effects ensure the consistency of the resulting low-energy physics across geometric transitions. The details depend on whether the transition point is approached along the Coulomb or the Higgs branch of the $\cN=2$ theory. Since in these setups, the vector multiplet moduli space/Coulomb branch does not receive any $g_s$ corrections, the effective action for the vector multiplet sector derived from perturbative string theory is exact. Still, the general properties of 4d $\cN=2$ supersymmetric theories dictate that a hypermultiplet must become massless at the origin of the Coulomb branch for the transition to the Higgs branch to be possible. As is well-known~\cite{Strominger:1995cz}, this hypermultiplet does not have a perturbative origin in string theory, but is instead accounted for by a wrapped D-brane. Thus, even though there are no non-perturbative corrections to the EFT couplings in the vector multiplet sector, the whole spectrum has to be completed by non-perturbative states to ensure consistency of the resulting physics across the geometric transition.

If instead we approach the transition point from the Higgs branch, the vector multiplet becoming massless at the transition locus arises as a perturbative string excitation. Unlike for the Coulomb branch, on the Higgs branch non-perturbative effects arising from D-brane instantons correct classical couplings derived from perturbative string theory. These ensure that the metric on the Higgs branch remains non-singular at the transition point~\cite{Ooguri:1996me,Saueressig:2007dr} as is required by supersymmetry.\footnote{Notice that~\cite{Ooguri:1996me} did not compute the instanton corrections directly in string theory. Instead, they imposed that supersymmetry requires the Higgs branch to have a HyperK\"ahler metric from which they derived the corrections to the string tree-level metric, which, subsequently, were \emph{interpreted} as instanton corrections.} Thus, non-perturbative effects are also crucial for a consistent realization of the transition from the Higgs branch to the Coulomb branch within string theory.\footnote{Similarly, as shown in~\cite{Marchesano:2019ifh,Baume:2019sry,Alvarez-Garcia:2021pxo} in certain asymptotic regimes quantum corrections to the hypermultiplet moduli space as computed in~\cite{RoblesLlana:2006is,RoblesLlana:2007ae,Alexandrov:2008gh,Alexandrov:2009zh,Alexandrov:2011va} are required to ensure consistency of the effective theory with the Distance~\cite{Ooguri:2006in} and Emergent String Conjecture~\cite{Lee:2019oct}.}  The above discussion illustrates that there are essentially two ways in which non-perturbative effects can ensure the consistency of the physical theory across geometric transitions: 
\begin{enumerate}
    \item[$i)$] The spectrum of light states contains states of non-perturbative origin. 
    \item[$ii)$] The couplings of the low-energy EFT are corrected by non-perturbative effects. 
\end{enumerate}
In the 4d $\cN=2$ theories arising from compactifications of Type~II string theory, whether the transition is approached along the Coulomb or Higgs branch determines which of the above non-perturbative effects plays a crucial role. This reflects in the factorization of the moduli space $\cM_{\cN=2} = \cM_{\rm VM}\times \cM_{\rm HM}$. Let us stress that from a field theory perspective, the transitions are nothing spectacular. It is just their realizations in string theory that require non-perturbative effects in $g_s$. 

The example of string-theoretic realizations of 4d $\cN=2$ theories illustrates that by taking into account non-perturbative effects, theories that seem to be disconnected at the perturbative level are in fact connected. In this way, the space of possible deformations is effectively enlarged at the non-perturbative level beyond what is accessible perturbatively. Based on the well-understood situation in 4d $\cN=2$ theories, a natural question is how non-perturbative effects alter realizations of 4d $\cN=1$ theories obtained from perturbative string theory. Compared to $\cN=2$ theories arising from string theory, much less is known about string compactifications preserving only $\cN=1$ supersymmetry in four dimensions. A main obstacle is that a similarly powerful technique, such as mirror symmetry, does not exist to describe $\cN=1$ string compactifications at the quantum level, though $(0,2)$ versions of mirror symmetry for the heterotic string have been studied in the literature~\cite{McOrist:2007kp,Kreuzer:2010ph}. Similarly, perturbative corrections to the K\"ahler potential are known perturbatively in Type IIB orientifold compactifications~\cite{Becker:2002nn,vonGersdorff:2005bf,Berg:2005ja,Berg:2007wt,Berg:2011ij,Berg:2014ama,Kim:2023sfs} and F-theory~\cite{Garcia-Etxebarria:2012bio,Grimm:2013gma,Grimm:2013bha,Minasian:2015bxa,Weissenbacher:2019mef,Klaewer:2020lfg,Cicoli:2021rub} whereas non-perturbative effects are mainly accessible only via duality to the heterotic string~\cite{Wiesner:2022qys,Cvetic:2024wsj}. In particular, geometric transitions in theories with minimal supersymmetry have been relatively unexplored, with conifold transitions in fourfolds being discussed in~\cite{Intriligator:2012ue} and extremal transitions in Calabi--Yau threefolds more recently in~\cite{Gendler:2022qof,Anderson:2022bpo}. The structure and properties of moduli spaces in 4d $\cN=1$ theories have also been investigated in the context of the swampland program. In particular, by analyzing infinite-distance limits in these moduli spaces, one can extract information about their geometric features as well as the spectrum of light states that emerge (see \cite{Font:2019cxq,Lanza:2020qmt,Lanza:2021udy,Cota:2022yjw,Marchesano:2022avb,Martucci:2022krl,Casas:2024ttx,Grieco:2025bjy}).

In this work, we focus on 4d $\cN=1$ theories arising as orientifolds of Type IIB string theory compactified on Calabi--Yau threefolds or, equivalently, F-theory on elliptically fibered Calabi--Yau fourfolds; for an introduction to the basics of F-theory, see~\cite{Weigand:2010wm,Weigand:2018rez}. In contrast to 4d theories with $\cN=2$ supersymmetry, the scalar field space of 4d $\cN=1$ theories does not factorize into decoupled sectors. Instead, if obtained by breaking half of the supersymmetries of a $\cN=2$ parent theory, the scalar fields of the hypermultiplets and vector multiplets of the parent theory surviving after supersymmetry breaking form part of $\cN=1$ chiral multiplets. Together, they form the chiral multiplet moduli space. As a consequence, this $\cN=1$ scalar field space combines features of both the vector and the hypermultiplet moduli space of the parent theory. On the one hand, similar to the 4d $\cN=2$ vector multiplet moduli space, the chiral multiplet moduli space is complex and K\"ahler. On the other hand, the chiral multiplet moduli space is subject to (non-)perturbative corrections in $g_s$, much like  $\cM_{\rm HM}$. Thus, non-perturbative physics leaves its imprints through corrections to the couplings of the low-energy EFT  as in $ii)$. Due to the absence of a factorization of the moduli space into two components, a natural expectation is that consistency of the 4d $\cN=1$ theory at the quantum level also requires light states that are of non-perturbative origin, as in $i)$. \newline

In this work, we investigate whether consistency of the 4d $\cN=1$ theory requires additional non-perturbative states to be present that are invisible from a perturbative Type IIB orientifold perspective. To maintain as much control as possible over the non-perturbative physics, we focus on subsectors of the full 4d $\cN=1$ theories that locally exhibit enhanced supersymmetry. We use this enhanced supersymmetry to infer the existence of additional states of non-perturbative origin in the 4d $\cN=1$ low-energy spectrum, which effectively enlarge the moduli space beyond what is accessible at the perturbative level. These subsectors of the whole 4d theory featuring enhanced supersymmetry can be viewed as the analogue of subsectors with locally enhanced supersymmetry in 6d $\cN=(1,0)$ compactifications of F-theory, as studied in~\cite{Morrison:1996na,Witten:1996qb}. In particular, in these cases the locally enhanced supersymmetry manifests as additional degrees of freedom that effectively enlarge the moduli space.

Exploiting the analogy to the previously studied six-dimensional setup, the main part of the analysis in this paper focuses on sectors with enhanced supersymmetry in the K\"ahler sector of Type IIB orientifold/F-theory compactifications to four dimensions. To that end, we consider F-theory compactified on elliptically fibered Calabi--Yau fourfolds and identify regions of the fourfold with local $SU(3)$ holonomy. These sectors are associated with a curve $C_0$ inside the base $B_3$ of the elliptically fibered fourfold that is not intersected by the anti-canonical divisor of $B_3$. In the Type IIB orientifold picture, these curves are hence not intersected by the O7-plane locus. To exploit the local supersymmetry enhancement in the vicinity of these curves, we focus on setups that can be decoupled from the gravitational sector of the theory, which globally only preserves four supercharges. To achieve this, we assume the curve $C_0$ to be shrinkable within $B_3$. A simple class of such shrinkable curves not intersecting the anti-canonical divisor are flop curves. 

Due to the enhanced supersymmetry, the small volume regime for $C_0$ can be identified with the vicinity of the origin of the Higgs branch of a 4d $\cN=2$ \emph{field theory} subsector of the theory. This implies that, in this regime, the light degrees of freedom associated with this field theory sector have to make up the field content of a massless $\cN=2$ hypermultiplet and a massive $U(1)$ vector multiplet. However, the dimensional reduction of Type IIB string theory on the CY orientifold yields a decoupled sector that only contains a massless $\cN=1$ chiral and a massive $\cN=1$ vector multiplet. As a central result of this work, we demonstrate that the missing degrees of freedom required to complete the $\cN=2$ multiplets arise once non-perturbative effects are taken into account. To that end, we consider the F-theory realization of the Type IIB orientifold. Crucially, in this formulation, there exist deformations that preserve the Calabi--Yau condition of the fourfold but change the topology of the base $B_3$. More precisely, we can consider the blow-up of the curve $C_0$ into an exceptional divisor $E$. The parameters associated with the blow-up provide the missing massless degrees of freedom to form a massless $\cN=2$ hypermultiplet. Importantly, these deformations are not visible from the perturbative Type IIB orientifold picture as the blow-up breaks the Calabi--Yau condition of the underlying Calabi--Yau threefold. From the perspective of the orientifold, these deformations are hence indeed of non-perturbative origin. 

Furthermore, by considering the massive degrees of freedom after the blow-up, including in particular states arising from wrapped D3-brane strings, we identify massive chiral multiplets that together with the massive $\cN=1$ vector multiplet provide the degrees of freedom of a massive $\cN=2$ vector multiplet. Given the non-perturbative description of the local field theory sector, a crucial question is whether this field theory sector is sensitive to the bulk gravitational theory, which overall preserves only four supercharges. Such a coupling to the bulk induces mild supersymmetry-breaking effects $\cN=2\to \cN=1$ in the local sectors. We can distinguish two such effects. The first type of supersymmetry breaking effects arises from the fact that the field theory emerges as a subsector of a gravitational theory. Geometrically, these effects are of topological nature and arise if $C_0$ is embedded in a compact Calabi--Yau fourfold and are absent only if a non-gravitational theory is considered, i.e., if the underlying Calabi--Yau fourfold is non-compact. Since these effects are topological, they do not depend on the relative coupling to gravity. Instead, the second type of supersymmetry breaking effects is non-topological and is sensitive to the gravitational coupling strength. In particular, these effects vanish in the limit where the overall volume is taken to infinity. This second type of supersymmetry breaking effects is induced by D3-brane instanton corrections. As we will see, for the decoupled field theory sector, these corrections render some states massive, generating an exponential mass hierarchy within the $\cN=2$ supermultiplets, thereby mildly breaking the locally enhanced supersymmetry.  \newline

For a second realization of local supersymmetry enhancement, we consider the complex structure moduli space of elliptically fibered Calabi--Yau fourfolds. To identify such sectors, we focus on supersymmetric F- and M-theory compactifications with non-zero $G_4$-flux. More precisely, we argue that supersymmetric flux vacua for which both the flux-induced superpotential and the F-term vanish can in particular be realized at loci in the complex structure moduli space where supersymmetry is enhanced. Our argument is based on considering the domain walls for which the superpotential gives the tension. In particular, along the supersymmetric loci, the domain wall becomes tensionless, such that we can realize the transition to the flux-less case at zero energy cost. Since the moduli space of the theory without flux is bigger than in the presence of flux, the existence of tensionless domain walls effectively enlarges the moduli space at the non-perturbative level. This provides a physical rationale for why one way of realizing supersymmetric flux vacua is at loci in the complex structure moduli space where the fourfold periods reduce to K3-like polynomial expressions as observed in~\cite{Grimm:2024fip}.

At first sight, the two main sectors with enhanced supersymmetry considered in this work seem unrelated. However, in a dual heterotic description, the two kinds of transitions that effectively enlarge the moduli space have a common origin. Concretely, via F-theory/heterotic duality~\cite{Morrison:1996ac,Morrison:1996pp}, certain F-theory compactifications on elliptically fibered Calabi--Yau fourfolds can be mapped to heterotic compactifications on elliptically fibered Calabi--Yau threefolds. Both types of transitions associated with sectors featuring enhanced supersymmetry correspond, on the dual heterotic side, to transitions involving the nucleation of spacetime-filling NS5-branes. Concretely, the nucleation of an NS5-brane along a two-cycle in the base of the heterotic Calabi--Yau threefold maps to the first type of transition realized in the K\"ahler moduli space. By contrast, wrapping an NS5-brane on the elliptic fiber of the heterotic Calabi--Yau threefold is equivalent to a transition corresponding to removing the $G_4$-flux in supersymmetric flux vacua. \newline

The rest of the paper is structured as follows: In Section~\ref {sec:6d}, we review the transition between two 6d $\cN=(1,0)$ theories arising from F-theory compactifications of Hirzebruch surfaces~\cite{Morrison:1996na}, which provides important guidance for the analysis in this paper. In Section \ref{sec:mojodojo} we study the analogue of such setups in 4d $\cN=1$ theories by analyzing sectors in the Kähler moduli space of F-theory compactifications that feature supersymmetry enhancement, focusing on simple flop curves and the non-perturbative completion of the spectrum of light states associated with such curves. In Section~\ref{sec:flux}, we then discuss an analogue of this transition in 4d flux compactifications of Type IIB/F-theory. In Section \ref{sec:heterotic}, we present the heterotic dual perspective, which ties together these two classes of transitions in a unified framework. In Section \ref{ssec:D3moduli} we extend our discussion to genuinely four-dimensional $\cN=1$ extremal transitions, and Section \ref{sec:discussion} we draw our conclusion.

\section{Enhanced Supersymmetry in 6d \texorpdfstring{$\cN=(1,0)$}{Lg} Theories}\label{sec:6d}

Consider a minimally supersymmetric theory of quantum gravity in six dimensions. Such 6d $\cN=(1,0)$ theories can be obtained, for example, by compactifying F-theory on an elliptically fibered Calabi--Yau threefold or as heterotic string theory compactified on a K3 surface with suitable gauge bundles. The low-energy approximation of these theories is a 6d $\cN=(1,0)$ supergravity theory. The massless degrees of freedom in this theory are arranged in supermultiplets for which the bosonic degrees of freedom are given by
\begin{itemize}
    \item One supergravity multiplet containing the graviton and a self-dual 2-form, 
    \item $n_H$ hypermultiplets containing four real scalars, 
    \item $n_T$ tensor multiplets containing a real scalar and an anti-self-dual two-form, 
    \item $n_V$ vector multiplets containing a gauge boson and no scalars. 
\end{itemize}
The matter content of the 6d $\cN=(1,0)$ theory of supergravity is constrained by anomaly cancellation conditions~\cite{Sagnotti:1992qw,Green:1984sg}. For example, cancellation of the gravitational anomaly requires 
\begin{equation}\label{eq:anomaly6d}
    273 + n_V = 29n_T +n_H\,. 
\end{equation}
A simple class of 6d $\cN=(1,0)$ theories corresponds to the case $n_T=1$. These can, for example, be obtained by perturbative\footnote{In this context, perturbative refers to the absence of spacetime-filling NS5-branes.} compactifications of the heterotic string on K3 or as F-theory compactified on elliptically fibered Calabi--Yau threefolds with base a Hirzebruch surface $\mathbb{F}_n$.

\subsection{Heterotic Perspective}
As shown in \cite{Morrison:1996na}, F-theory compactifications with base $\mathbb{F}_n$ are dual to the heterotic $E_8\times E_8$ string on K3 with instanton embedding $(12-n, 12+n)$.

\[
\text{F-theory on CY3} 
\;\;\xleftrightarrow{\hspace{3cm}}\;\;
\text{Heterotic $E_8 \times E_8$ on K3}
\]
\[\hspace{0em}
\mathcal{E} \;\longrightarrow\; X_3 \;\longrightarrow\; \mathbb{F}_n 
\qquad\qquad \hspace{4.8em}
24 \;=\; (12-n) + (12+n)
\]

In this work, we are particularly interested in the two models corresponding to $n=0$ and $n=2$. In the following, we refer to the theory labelled by $n$ as $\mathbb{T}_n$. As shown in \cite{Morrison:1996na} (see also \cite{Aldazabal:1996fm}), these two models are in fact dual to each other and share a common moduli space with $n_H=244$.  Therefore, starting from $\mathbb{T}_2$ there must be a deformation that brings us from $\mathbb{T}_2\to \mathbb{T}_0$. In the heterotic formulation, such a transformation looks rather drastic: after all, we change the topology of the gauge bundle in both $E_8$ gauge factors. Equivalently, taking the perspective of heterotic M-theory \cite{Horava:1995qa,Horava:1996ma}, we change the flux in the 9-branes at the end of the $S^1/\mathbb{Z}_2$ interval. To see this, recall that for M-theory on $S^1/\mathbb{Z}_2$ two $E_8$ gauge theories are realized in the 9-branes at the end of the interval. Depending on the spacetime and gauge background, there is $G_4$-flux localized in these branes. Suppose we parametrize the interval of length $\rho$ by a coordinate $y\in [-\rho/2,\rho/2]$ and denote the 9-branes localized at $y=\pm \rho/2$ by $\textbf{9} _\pm$. If, furthermore, the background gauge field strength in the 9-branes is denoted by $F^{\pm}$, the $G_4$-flux charge in the 9-branes is given by 
\begin{align}
      Q_{G_4}^{\pm} = \Tr(F^\pm\wedge F^\pm) -\frac12 \tr R\wedge R\,. 
\end{align}
For the M-theory uplift of the heterotic string compactified on K3 with instanton embedding $(12-n,12+n)$ the $G_4$ flux is simply 
\begin{equation}
    Q_{G_4}^\pm = \pm n\,. 
\end{equation}
From the perspective of M-theory, the transition $\mathbb{T}_2 \to \mathbb{T}_0$ can thus be viewed as removing the flux localized in the 9-branes entirely. As explained in \cite{Seiberg:1996vs}, this can be achieved by nucleating two M5-branes from $\textbf{9}_+$ and absorbing them in $\textbf{9}_-$, which involves traversing various small instanton transitions~\cite{Witten:1995gx}. This process is hence not smooth. Still, on general grounds, one expects that there exists a smooth transition between $\mathbb{T}_2$ and $\mathbb{T}_0$. However, from the perturbative heterotic perspective, such a smooth transition would be associated with a domain wall with tension of order the Planck scale and hence not describable within the validity of the effective field theory derived from the perturbative heterotic string.
\subsection{F-theory Perspective}
Instead, the dual F-theory perspective reveals that a smooth transition from $\mathbb{T}_2\to \mathbb{T}_0$ has to be possible dynamically within (a dual) effective field theory description. In particular, in this formulation, the transition merely corresponds to a deformation in the (extended) moduli space of the theory. Interestingly, this means not only that the transition is understood within effective field theory, but also that it is possible at zero energy. 

To see this, let us briefly review the concrete construction of the $\mathbb{T}_2$ model as in~\cite{Morrison:1996na}. Therefore, consider the degree 24 hypersurface in $\mathbb{P}_{1,1,2,8,12}$ which is of the general form 
\begin{equation}
    x_1^{24}+ x_2^{24} + x_3^{12} +x_4^3 +x_5^2 + \sum_i \psi^i P_i^{(24)}(x_1,\dots,x_5)=0\,,
\end{equation}
where the $P_i^{(24)}$ are polynomials of degree $24$ of the coordinates $x_i$. There are 242 such polynomials corresponding to 242 \emph{polynomial} complex structure deformations $\psi^{i=1,\dots, 242}$. After resolving all orbifold singularities, the hypersurface Calabi--Yau threefold can be viewed as a smooth elliptic fibration over $\mathbb{F}_2$. Thus, compactifying F-theory on $\mathbb{P}_{1,1,2,8,12}[24]$ yields a six-dimensional theory with $n_T=1$, $n_V=0$. The complex structure deformations $\psi^i$ and the overall volume modulus of $\mathbb{F}_2$ give rise to $243$ hypermultiplets. However, this cannot be the final answer since with this amount of hypermultiplets the RHS of \eqref{eq:anomaly6d} is off by one. This would suggest that F-theory on ${\mathbb{T}_2}$ has to be anomalous. To cancel the gravitational anomaly, we are hence missing one hypermultiplet that cannot be realized as a polynomial deformation of $X_3^{\mathbb{T}_2}= \mathbb{P}_{1,1,2,8,12}[24]$. As shown in \cite{Morrison:1996na}, this additional hypermultiplet corresponds to a non-polynomial deformation that becomes manifest when describing the Calabi--Yau threefolds as a complete intersection instead of a hypersurface. 

From the perspective of the base, this non-polynomial deformation can be viewed as follows~\cite{Morrison:1996na}: Blowing down the $(-2)$-curve in $\mathbb{F}_2$, the base of the elliptic Calabi--Yau threefold can be represented as the hypersurface
\begin{equation}
   \{ y_1 y_2  + y_3^2 =0\} \subset \mathbb{P}^3
\end{equation}
where $(y_1,y_2,y_3,y_4)$ are the coordinates of $\mathbb{P}^3$. The non-polynomial deformation now corresponds to the deformation 
\begin{equation}
    y_1 y_2 + y_3^2 = \psi_{\rm n.p.} y_4^2 \,. 
\end{equation}
For $\psi_{\rm n.p.}\neq 0$ the surface is topologically $\mathbb{F}_0$. In other words, the theory $\mathbb{T}_2$ is realized at a special locus $\cM_{\mathbb{T}_2}$ inside the hypermultiplet moduli space of the $\mathbb{T}_0$ theory. Let us denote by $X_3^{\mathbb{T}_0}$ the complete intersection Calabi--Yau threefold describing the general $\mathbb{T}_0$ theory. Then we have 
\begin{equation}
    \Delta h^{2,1} \equiv h^{2,1}\left(X_3^{\mathbb{T}_0}\right) -  h^{2,1}_{\rm pol.}\left(X_3^{\mathbb{T}_2}\right) \ .
\end{equation} 
The modulus $\psi_{\rm n.p.}$ (together with its axionic partners) comprises the missing hypermultiplet to satisfy the anomaly cancellation condition~\eqref{eq:anomaly6d}. In the hypermultiplet moduli space of $\mathbb{T}_0$, the locus $\cM_{\mathbb{T}_2}$ is thus a $\text{codim}_{\mathbb{R}}=4$ locus. 

The F-theory discussion illustrates that turning on the more general deformation $\psi_{\rm n.p.}$ allows for a smooth transition from $\mathbb{T}_2$ to $\mathbb{T}_0$. This demonstrates that the transition between these two theories is possible dynamically at zero energy. We can also understand the reverse transition $\mathbb{T}_0\to \mathbb{T}_2$. Therefore we consider the moduli space $\cM_{\mathbb{T}_0}$. At generic points in this moduli space, the spectrum of BPS strings is generated by D3-branes wrapping the two $\mathbb{P}^1$s in $\mathbb{F}_0=\mathbb{P}^1_f\times \mathbb{P}^1_g$. Accordingly, there are two strings which we denote by $\mathtt{S}_f$ and $\mathtt{S}_g$. At generic points in $\cM_{\mathbb{T}_0}$, the curve $f-g$ is not holomorphic, such that a D3-brane wrapped on it does not yield a BPS string. This is different along $\cM_{\mathbb{T}_2}$. Here an additional holomorphic curve $h=f-g$ arises that yields a BPS string $\mathtt{S}_h$ upon wrapping a D3-brane on it. From the perspective of the base $\mathbb{F}_2$, the curve $h$ is simply the $(-2)$-curve that can be identified as the section $\CZ_-$ of $\mathbb{F}_2$. The transition $\mathbb{T}_0 \to \mathbb{T}_2$ can be viewed as nucleating a pair of strings $\pm\mathtt{S}_g$ that form bound states with $\mathtt{S}_f$ only along $\mathcal{M}_{\mathbb{T}_2}$. The second bound state $\mathtt{S}_{f+g}$ then corresponds to a D3-brane on the $\CZ_+$ section of $\mathbb{F}_2$. 

The curve $h$ has the special property that it does not intersect the anti-canonical class of $\mathbb{F}_2$. To see this, we note that 
\begin{equation}
    \bar{K}_{\mathbb{F}_2} = 2h+4f= 2f+2g =\bar{K}_{\mathbb{F}_0} \,.
\end{equation}
Together with the intersection ring 
\begin{equation}
    h\cdot_{\mathbb{F}_2} h=-2\,,\quad f\cdot_{\mathbb{F}_2} f=0\,,\quad f\cdot_{\mathbb{F}_2} h=1\,,
\end{equation}
this indeed implies $\bar{K}_{\mathbb{F}_2}\cdot_{\mathbb{F}_2} h= 0$. From the F-theory perspective, the curve thus does not intersect the O7-plane locus, such that in the vicinity of $h$ we can locally realize enhanced supersymmetry.\footnote{This local enhancement can also be understood through the connection between F-theory and Type IIB string theory. In the neighborhood of $h$, the geometry (although $\mathbb{F}_2$ itself is not) locally resembles a hyper-Kähler manifold. As a result, the Type IIB string coupling does not vary over this region. Since perturbative Type IIB string theory possesses twice the amount of supersymmetry compared to the heterotic string, this explains the local $\mathcal{N}=2$ enhancement.} Accordingly, the string $\mathtt{S}_h$ can be identified as an $\cN=(2,0)$ SCFT string. This SCFT string couples to an anti-self dual tensor field that, from the perspective of the bulk $\cN=(1,0)$ supersymmetry, resides in an $\cN=(1,0)$ tensor multiplet. This tensor multiplet contains a real scalar that, in the small volume limit for $h$, can be identified with the volume of $h$. However, due to the locally enhanced supersymmetry, the tensor should in fact be part of a $\cN=(2,0)$ matter multiplet, which in the $\cN=(1,0)$ language consists of a tensor and a hypermultiplet. Thus, the locally enhanced supersymmetry dictates that the tensor multiplet has to pair up with one of the hypermultiplets to form an $\cN=(2,0)$ matter multiplet. As explained in~\cite{Witten:1996qb}, this hypermultiplet is the one associated with the non-polynomial deformation $\psi_{\rm n.p.}$ invisible from the hypersurface equation in $\mathbb{P}_{1,1,2,8,12}$. 

Compared with the heterotic/M-theory realization, one can understand the transition as a domain wall between two EFTs. From the point of view of heterotic/M-theory, the necessary domain wall would have to brute force extract NS5-branes from one 9-brane location and place them on the other side, which would require small instanton transitions. However, from the F-theory perspective, the transition between the theories is smooth and occurs at zero energy cost, since the two EFTs share a common moduli space.

In summary, the naive moduli space $\cM_{\mathbb{T}_2}$ is embedded in the larger one, given by $\cM_{\mathbb{T}_0}$. The key features of this 6d setup can be summarized in the following way: 
\begin{enumerate}
    \item The codimension of the naive moduli space is  \begin{equation} \text{codim}_\mathbb{R}(\cM_{\mathbb{T}_2}\subset \cM_{\mathbb{T}_0}) =4\,,
    \end{equation} and the directions transverse to $\cM_{\mathbb{T}_2}\subset \cM_{\mathbb{T}_0}$ are the scalar components of the hypermultiplet associated with $\psi_{\rm n.p.}$.
    \item Along $\cM_{\mathbb{T}_2}\subset \cM_{\mathbb{T}_0}$ there is an additional holomorphic curve $h=f-g$. 
    \item The curve $h$ does not intersect $\bar{K}_{\mathbb{F}_2}$ and thus corresponds to a subsector with enhanced supersymmetry. 
    \item The degrees of freedom becoming light at $\text{vol}(h)=0$ correspond to a 6d $\cN=(2,0)$ SCFT and thus also respect the enhanced supersymmetry.
    \item In the heterotic/M-theory formulation, the transition $\mathbb{T}_2 \to \mathbb{T}_0$ corresponds to removing $G_4$-flux from the 9-branes at the end of the M-theory interval. 
\end{enumerate}

The relation between these 6d theories is closely tied to the enhanced supersymmetry in $\mathbb{T}_2$ and serves as a warm-up for our discussion of similar setups in 4d $\cN=1$ realizations of F-theory that feature sectors with enhanced supersymmetry. Such setups are the main interest of the following sections.

\section{SUSY Enhancement in the K\"ahler Moduli Space}\label{sec:mojodojo}

We now shift our attention to four-dimensional theories with minimal supersymmetry realized as compactifications of F-theory on elliptically fibered Calabi--Yau fourfolds. In this context, we are interested in identifying smooth transitions analogous to the $\mathbb{T}_2 \to \mathbb{T}_0$ transition discussed in the previous section. The transition described in section~\ref{sec:6d} can either be interpreted as a transition between two realizations of heterotic/M-theory differing in the flux quanta, or in an F-theory context as a smooth transition between two different bases of an elliptically fibered Calabi--Yau threefold that share a common moduli space. 
In this section, we aim to identify the 4d $\cN=1$ analogue of the second point of view.
To that end, we will focus on subsectors of the 4d $\cN=1$ theory of gravity featuring enhanced supersymmetry characterized by a subsector of the geometry with local $SU(3)\subset SU(4)$ holonomy in the full fourfold. In other words, we study subsectors of the Calabi--Yau fourfold in F-theory that locally look like a trivial elliptic fibration over a Calabi--Yau threefold without an orientifold projection. We will exploit the enhanced supersymmetry associated with these subsectors to guide us in uncovering non-perturbative effects in small volume regimes of F-theory.

\subsection{Flop Curves in F-theory}

Consider F-theory compactified on an elliptically fibered Calabi--Yau $X_4$ fourfold with base $B_3$, 
\begin{equation}\label{fourfoldfibration}
    T^2 \hookrightarrow X_4 \to B_3\,. 
\end{equation}
By analogy to the six-dimensional case discussed in section~\ref{sec:6d}, we consider a geometric subsector featuring enhanced supersymmetry associated with a curve $C_0$ in $B_3$ satisfying
\begin{equation}
    \bar{K}_{B_3} \cdot C_0 =0\,,
\end{equation}
where $\bar{K}_{B_3}$ is the anticanonical divisor of $B_3$. We further assume that the curve $C_0$ is shrinkable such that it describes a subsector that can be decoupled from gravity. We thus impose that $C_0$ is effective and $g(C_0)=0$. We can write $C_0=D_1\cdot D_2$ with $D_{1,2}$ effective divisors on $B_3$. Adjunction yields
\begin{equation}
    2-2g(C_0) = (\bar{K}_{B_3} - D_1 -D_2)\cdot D_1 \cdot D_2\,. 
\end{equation}
There are three possibilities for the normal bundle of the curve $C_0$ inside $B_3$, namely
\begin{align}
    \mathcal{N}_{C_0/B_3} = \left\{\begin{matrix}
        \mathcal{O}(-1)\oplus O(-1)\,, \\ 
        \mathcal{O}(-2) \oplus \mathcal O(0) \,,\\
        \mathcal{O}(-3) \oplus \mathcal{O}(1)\,. 
    \end{matrix}\right.
\end{align}
In the following, we focus on the first case and consider an effective curve $C_0$ with normal bundle 
\begin{equation}\label{normalbundleF}
\cN_{C_0/B_3}= \mathcal{O}(-1)\oplus \mathcal{O}(-1)\,.
\end{equation} 
The curve $C_0$ then corresponds to an Atiyah flop curve~\cite{atiyah1958}. In other words, we can shrink the curve $C_0$ to a point without shrinking any divisor of $B_3$ or $B_3$ itself to a lower-dimensional (sub-)manifold. Flops in the base of elliptically fibered Calabi--Yau fourfolds have previously played a critical role in~\cite{Cota:2022yjw}. Furthermore, the physics associated with flop transitions in the fiber of such fourfolds has been investigated, e.g., in~\cite{Hayashi:2013lra}.

In the following, it is instructive to also consider such curves in Sen's orientifold limit~\cite{Sen:1997gv} for the full fourfold~\eqref{fourfoldfibration}. In this limit, we can identify a Calabi--Yau threefold $X_3$ that is the double cover of $B_3$. Since the curve $C_0$ does not intersect the anticanonical class of $B_3$ it lifts to a curve $\tilde{C}_0$ in $X_3$ with normal bundle 
\begin{equation}\label{normalbundletildeC}
   \cN_{\tilde{C}_0/X_3} = \cO(-1)\oplus \cO(-1)\,.  
\end{equation}
Thus, the Calabi--Yau threefold $X_3$ also contains a simple flop curve. Before orientifolding, Type IIB string theory compactified on $X_3$ contains a hypermultiplet for which the scalar degrees of freedom are given by
\begin{equation}\label{TypeIIBfields}
    \tilde t_0 = \int_{\tilde{C}_0} J_{X_3} \,,\quad  \tilde b_2^{(0)} = \int_{\tilde C_0} B_2\,,\quad \tilde c_2^{(0)} = \int_{\tilde C_0} C_2 \,, \quad \tilde c_4^{(0)}= \left(\int_{\tilde C_0} C_4\right)^\vee\,,
\end{equation}
where $J_{X_3}$ is the K\"ahler form on $X_3$ that measures lengths in units of the (inverse) Planck scale $M_{10}=g_s^{-1/4}M_s$ with $M_s$ the Type IIB string scale. Moreover, $B_2/C_2$ are the Type IIB NSNS/RR 2-forms, and $C_{4}$ is the RR 4-form of Type IIB string theory, and $()^\vee$ dualizes a 2-form into an axion in four dimensions. 

The orientifold action projects out half of the degrees of freedom of the above hypermultiplet. Since the curve $\tilde{C}_0\subset X_3$ maps to the curve $C_0\subset B_3$ in the base of $X_4$, the curve $C_0$ lies in $H_{2}^{ +}$ such that the orientifold action projects out $\tilde b_2^{(0)}$ and $\tilde c_2^{(0)}$ as dynamical degrees of freedom. For a curve $C_0$ not intersecting the anticanonical class of $B_3$, anomaly cancellation requires
\begin{equation}
\langle b_2^{(0)}\rangle =\langle c_2^{(0)}\rangle =0\,.
\end{equation}
The remaining degrees of freedom of the 4d $\cN=2$ hypermultiplet then map to massless chiral multiplets in the 4d $\cN=1$ theory that remain as classically massless scalar degrees of freedom also in the F-theory description
\begin{equation}
    \tilde t_0  \mapsto t_0\equiv \int_{C_0} J_{B_3} \,,\qquad \tilde c_4^{(0)}\mapsto c_4^{(0)}\equiv  \left(\int_{} C_4\right)^\vee\,.
\end{equation} 

As $C_0$ does not intersect the orientifold locus, it locally sees enhanced supersymmetry.\footnote{The difference between curves intersecting the orientifold locus and those that do not has also been discussed in~\cite{Demirtas:2021nlu}.} In fact, the vicinity of $C_0$ in $B_3$ is locally a Calabi--Yau threefold, and the $\cN=2\to \cN=1$ supersymmetry breaking effects in the sector associated with $C_0$ can be made arbitrarily weak by pushing the O7-planes far away from $C_0$. This is, of course, only possible since $C_0$ is not intersected by the divisor $[O7]$ hosting the O7-planes. Since $C_0$ is shrinkable, we can decouple the sector associated with $C_0$ from the full gravitational theory and thus obtain an $\cN=2$ field theory subsector in a theory of gravity that globally only preserves $\cN=1$ supersymmetry.

In the following, we focus on the subsector of the theory associated with the curve $C_0$ that can be entirely decoupled from gravity. For this reason, we can safely assume that the threefold $B_3$ containing the flop curve $C_0$ is non-compact. For most of this section, we analyze the small volume regime for $C_0$ in this local setup with non-compact base and delegate the discussion of global effects on compact bases $B_3$ to section~\ref{ssec:global}.

\subsubsection{Small volume limit in $\cN=2$ theories} 
 To discuss the small volume regime of the $\cN=2$ subsector embedded in the global $\cN=1$ theory, we first review the physics of the small volume regime of flop curves in Calabi--Yau compactifications of Type IIB string theory. In Type IIB compactifications on a Calabi--Yau threefold $X_3$, the small volume regime for a curve $\tilde C_0$ with normal bundle as in~\eqref{normalbundletildeC} is well understood, see \cite{Strominger:1995cz,Greene:1995hu,Greene:1996dh} in the context of conifold transitions in 4d $\cN=2$ theories. 
 
 Consider the simplest scenario involving two homologous curves, $\tilde C_0$ and $\tilde C_0^h$. The low-energy physics resulting from dimensional reduction is equivalent to the case with a single curve $\tilde C_0$, with the crucial difference that there exists a three-chain $\Sigma_3$ satisfying \begin{equation} 
 \partial \Sigma_3 =\tilde  C_0 -\tilde C_0^h\,. 
 \end{equation}
 A massive $U(1)$ gauge field is supported along this chain. More precisely, the massive vector field arises from the dimensional reduction of the RR four-form $C_4$ over $\Sigma_3$. The limit of vanishing volume of the curves $\tilde C_0^{(h)}$ corresponds to the origin of an $\mathcal{N}=2$ Higgs branch, where a massive $\mathcal{N}=2$ $U(1)$ vector multiplet becomes massless and splits into a massless vector multiplet and a massless, charged hypermultiplet \cite{Greene:1995hu,Greene:1996dh}. The massless vector multiplet signals that the $U(1)$ gauge symmetry is restored. In general, the bosonic degrees of freedom of a given supermultiplet can be decomposed into $n_S$ scalar and $n_V$ vector components which we summarize as $(n_S, n_V)$. In this notation, the split described above can be represented as
\begin{equation}
    (5,3)_{V, m\neq 0} \longrightarrow (2,2)_{V,m=0} \,+ (4,0)_H\,, \label{dof vector}
\end{equation}
where the index distinguishes between vector ($V$) or hypermultiplets ($H$) of the 4d $\cN=2$ theory. The degrees of freedom that become light at the origin of the Higgs branch and split as in \eqref{dof vector} have different origins depending on whether we approach the conifold point from the Higgs or from the Coulomb branch.
\begin{figure}[t!]
    \centering
    \includegraphics[width=0.75\linewidth]{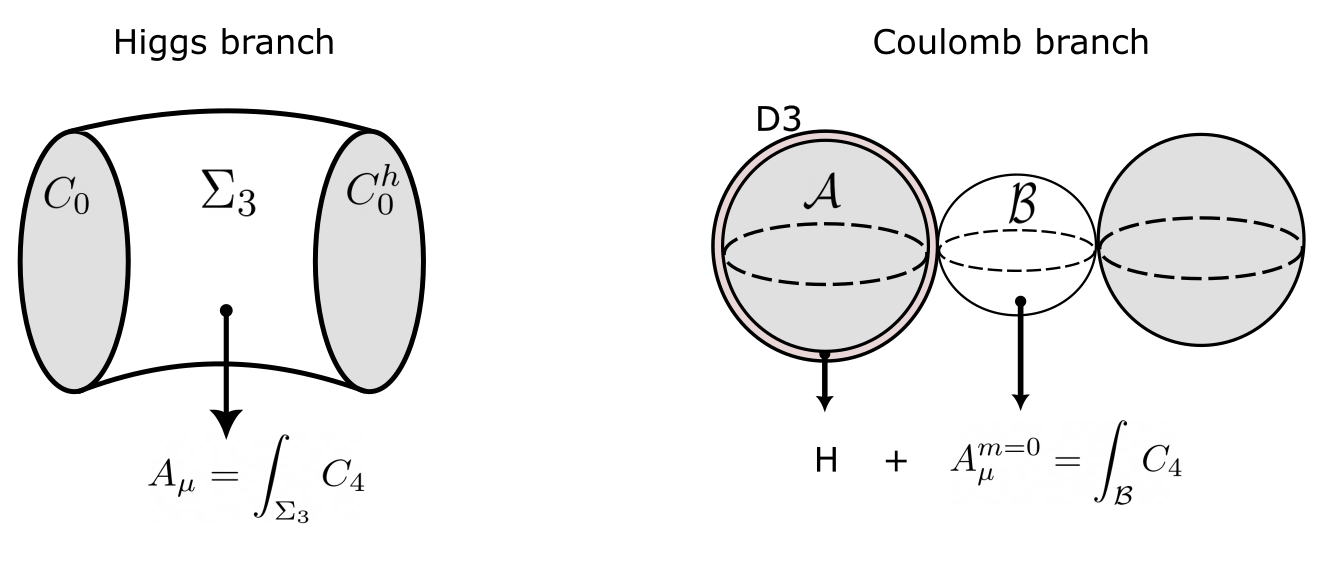}
    \caption{The conifold singularity of the 4d $\mathcal{N}=2$ theory separates the Higgs from the Coulomb branch. The Higgs branch is parameterized by the volume of $\tilde C_0$, while the Coulomb branch is parameterized  by the volume of $\mathcal{A}$. In the Higgs branch, a massive photon arises by reducing $C_4$ over the chain $\Sigma_3$. In the Coulomb branch, the same amount of bosonic degrees of freedom arises from a charged hypermultiplet corresponding to D3-brane wrapping $\mathcal{A}$, plus the dimensional reduction of $C_4$ over $\mathcal{B}$, yielding a massless gauge boson.}
    \label{fig:placeholder}
\end{figure}
\begin{itemize}
 \item From the Higgs branch perspective, the massive vector multiplet arises from dimensionally reducing the Type IIB supergravity action. More concretely, the vector degrees of freedom originate from reducing the four-form potential $C_4$ over the three-chain $\Sigma_3$ and are hence perturbative states of Type IIB string theory. Approaching the origin of the Higgs branch, the three-chain $\Sigma_3$ then transforms into a three-cycle $\mathcal{B}$. Let us denote its magnetic dual by $\mathcal{A}$ (i.e., $\mathcal{A} \cdot \mathcal{B} = 1$). The classical singularity arising at this point in moduli space is resolved by summing over $[p,q]$-string instantons wrapping $\tilde{C}_0$, and a three-cycle $\mathcal{A}$ emerges \cite{Greene:1995hu,Ooguri:1996me}. At this point, we can transition onto the Coulomb branch \cite{Strominger:1995cz} along which the origin of the massless fields is different. 
\item From the perspective of the Coulomb branch, the degrees of freedom making up the massive vector multiplet in \eqref{dof vector} arise from a massless vector multiplet (the first term in \eqref{dof vector}) obtained from the reduction of $C_4$ over $\mathcal{B}$ and the hypermultiplet originating from a D3-brane wrapping $\mathcal{A}$ \cite{Strominger:1995cz,Greene:1995hu}. In particular, the D3 hypermultiplet becomes the longitudinal modes of the massless vector multiplet by a Higgs mechanism process.  
\end{itemize} 
As one can see, the massive vector multiplet has two different origins depending on the perspective, i.e., whether we approach the conifold along the Coulomb or Higgs branch. Along the Higgs branch, the states in the massive vector multiplet arise as perturbative string states. In contrast, along the Coulomb branch, non-perturbative D3-brane states have to be included to complete the massive $\cN=2$ vector multiplet. This observation illustrates a duality between the closed string spectrum and the non-perturbative spectrum arising from D-branes \cite{Greene:1995hu}. 

\subsubsection{Type IIB Orientifold Perspective}
The physics in the small volume limit of the curve $\tilde{C}_0$ in an actual 4d $\cN=2$ theory will underlie the discussion that follows. Let us first discuss the situation after orientifolding from the perturbative Type IIB perspective, before moving on to the full F-theory discussion. Breaking half of the supersymmetry globally projects out half of the bosonic degrees of freedom. Here we focus on the case that after the orientifold projection a massive $\cN=1$ vector multiplet remains in the spectrum for which the bosonic field content can be represented as
\begin{equation}
    \text{massive }\cN=1\text{ vector:} \quad (n_S,n_V)=(1,3)\,. 
\end{equation}
In terms of massless $\cN=1$ multiplets this can be viewed as the scalar field content of a massless vector and a chiral $\cN=1$ multiplet, i.e.,
\begin{equation}
    (1,3)_{V, m\neq0}\simeq (0,2)_{V,m=0}\; + (2,0)_C\,.
\end{equation}
In other words, after orientifolding, the massive 4d $\cN=2$ vector multiplet has lost the degrees of freedom of a 4d $\cN=2$ hypermultiplet, or equivalently, two 4d $\cN=1$ chiral multiplets. 

In the vicinity of $\text{vol}(C_0)=0$ of the orientifolded theory, the light, perturbative field content is given by a massless chiral multiplet with scalar degrees of freedom $t_0$ and $c_4^{(0)}$ and a massive $\cN=1$ vector multiplet. As the curve $C_0$ is not intersected by $[O7]$, the theory in the vicinity $\text{vol}(C_0)=0$ has enhanced supersymmetry. This means that the light degrees of freedom should arrange into full $\cN=2$ multiplets. The origin of these additional degrees of freedom is not obvious from the Type IIB orientifold perspective since, as we just saw, the orientifold generically projects out half of the light degrees of freedom of the \emph{perturbative} Type IIB theory. Since the perturbative theory does not provide the necessary degrees of freedom, these have to be of non-perturbative origin. 

The situation so far is reminiscent of the 6d compactification of F-theory on the degree-24 hypersurface in $\mathbb{P}_{1,1,2,8,12}$ \cite{Morrison:1996na}. Here, the additional degrees of freedom cannot be detected directly from the hypersurface equation but correspond to \emph{non-polynomial deformations} of the hypersurface which are visible in the complete intersection realization of this model. 
In our 4d setup, the analogue of non-polynomial deformations of the hypersurface Calabi--Yau threefold corresponds to deformations of $B_3$ that do not preserve the Calabi--Yau condition of the original threefold $X_3$. They are hence not visible from the perturbative Type IIB compactification and become visible only in F-theory, as we discuss in the following section. 

\subsection{Non-Perturbative Phase Structure}\label{ssec:additionaldeformation}
Returning to the full F-theory setup, we consider again the curve $C_0$ in the non-compact base $B_3$ with normal bundle as in~\eqref{normalbundleF}. Flopping this curve into $-C_0$ leads to a birationally equivalent base, $B_3'$. The locus $\{t_0=0\}$ is classically a real codimension-one singularity in the moduli space. However, since the K\"ahler moduli space is complex, this singularity has to be resolved such that it is possible to ``move around'' the singular locus. This can be done by changing the vev of the axionic partner of $t_0$, $c_4^{(0)}$. However, due to the enhanced supersymmetry, the moduli space in the vicinity of $\{t_0=0\}$ has to resemble the origin of an $\cN=2$ field-theoretic Higgs branch. For this reason, there have to be two additional real directions to form a complete $\cN=2$ hypermultiplet. In the following, we identify the origin of these extra degrees of freedom in the F-theory realization of the theory.

\subsubsection{Birational Factorization of Flop}
\begin{figure}[t!]
    \centering
\includegraphics[width=0.6\linewidth]{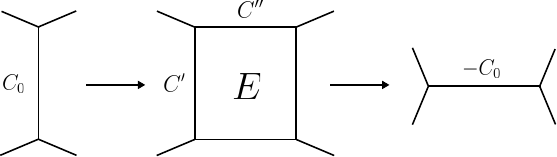}
    \caption{Schematic representation of the birational factorization of a flop transition. The left diagram corresponds to the 4-fold $X_4$ containing the flop curve $C_0$, whereas the right one corresponds to $\tilde X_4$ after a flop transition. The diagram in the middle shows the 4-fold $\hat X_4$ for which the curve $C_0$ has been blown up into an extra exceptional divisor $E$.}
    \label{fig: flop transition}
\end{figure}

Let us recall that the orientifold projection removes $\tilde b_2^{(0)}$ and $\tilde c_2^{(0)}$ from the spectrum, which, in the Type IIB compactification on $X_3$, complete $(\tilde t_0, \tilde c_4^{(0)})$ to an $\cN=2$ hypermultiplet. To identify the missing scalar field directions, we therefore have to consider a deformation that is invisible to the original Calabi--Yau threefold. A natural candidate for such a deformation corresponds to the birational factorization of the flop~\cite{Reid:1983}: For a simple local model of a flop transition, consider the singular quadric $Q$ in $\mathbb{P}^3$ given by 
\begin{equation}
    Q= \{x_1 x_2 + x_3^2 = 0\} \subset \mathbb{P}^3\,. 
\end{equation}
Blowing up the origin, one obtains a birational morphism $W\rightarrow Q$ with exceptional divisor $E\simeq \mathbb{P}^1\times \mathbb{P}^1$. The exceptional divisor $E$ can be partially contracted in two ways, $W\rightarrow X$ and $W\rightarrow Y$, where the two manifolds $X$ and $Y$ are related via a flop transition passing through the singular cone $Q$. The series of blow-ups and contractions described by $X\leftarrow W \rightarrow Y$ is the birational factorization of the flop. A simple flop in a compact threefold can locally be viewed as a variant of this simple quadric example. The birational factorization of the flop transition from $C_0$ to $-C_0$ is shown schematically in Fig.~\ref{fig: flop transition}. 

Let us denote by $\pi: X_4\rightarrow B_3$ the fourfold containing the flop curve $C_0$ and by $\check{\pi}:\check{X}_4\rightarrow \check{B}_3$ the fourfold after the flop transition. In addition, we can consider the fourfold $\widehat{\pi}:\widehat{X}_4 \rightarrow \widehat{B}_3$ for which $C_0$ is blown up as in the birational factorization of the flop. Compared to $B_3$ and $\check{B}_3$ the base $\widehat{B}_3$ contains an extra exceptional divisor $E\subset B_3$. Let us, for the moment, focus on the bases $B_3,\check{B}_3, \widehat{B}_3$ and ignore the additional elliptic fibrations over them. For these bases, consider the birational maps $\phi:\widehat{B}_3\to B_3$ and $\check{\phi}:\widehat{B}_3 \to \check{B}_3$ and denote by $C'$ and $C''$ the two $\mathbb{P}^1$ factors of $E\simeq \mathbb{F}_0$. We notice that the two contractions $\phi$ and $\check\phi$ map 
\begin{equation}\label{C0Cprime}
  C_0= \phi_*(C') = -\check{\phi}_*(C'')\,,
\end{equation} 
such that the proper transform of the curve $C_0 \in H_2(B_3)$ lies in the class $[C'-C''] \in H_2(\widehat{B}_3)$. On the other hand, the anti-canonical divisors of $B_3$ and $\widehat{B}_3$ are related via 
\begin{equation}\label{shiftKbar}
\bar{K}(\widehat{B}_3)= \phi^*(\bar{K}(B_3)) - E\,. 
\end{equation}
Given the non-trivial transformation of the anti-canonical class upon blowing up $C_0$, it is clear that the following discussion is invisible from the perspective of Type IIB orientifolds. In this description, one starts from a Calabi--Yau threefold $X_3$ that contains the flop curve $\tilde C_0$. However, as implied by \eqref{shiftKbar}, the birational factorization associated with the flop breaks the Calabi--Yau condition. Consequently, the deformation corresponding to this birational transition cannot be captured within the perturbative framework of Type IIB compactifications on Calabi--Yau orientifolds.

\subsubsection{Geometric Degrees of Freedom after Blow-up}
Blowing up the curve $C_0$ introduces an additional chiral multiplet in the theory, for which the scalar degrees of freedom are given by
\begin{equation}
    T_E = \frac{1}{2} \int_E J_{\widehat {B}_3}^2 + i\int_E C_4\,,
\end{equation}
where $J_{\widehat{B}_3}$ is the K\"ahler form on the blown-up base $\widehat{B}_3$ again measuring length in terms of the (inverse) 10d Planck scale $M_{10}$. 
At the purely geometric level, the moduli space of the original model corresponding to the threefolds $B_3$ and $\check{B}_3$ can be identified as the locus 
\begin{equation}
    \cM_{B_3}\cup\cM_{\check{B}_3} \stackrel{\rm cl.}{=}\{T_E = 0\} \subset \cM_{\widehat{B}_3}\,. 
\end{equation}
However, from the perspective of F-theory on an elliptic fibration $T^2\hookrightarrow \widehat{X}_4\rightarrow \widehat{B}_3$, this locus is in the deep quantum regime as it corresponds to the limit where (at least classically) an effective divisor shrinks to zero size. We get back to the properties of this quantum regime below, but for the moment, we stick to the geometric description. 

For compact models, a blow-up of a curve leads to a change in the Euler characteristic of the fourfold and thus of the geometry-induced D3-brane tadpole that has to be canceled either by $G_4$-flux or by spacetime-filling D3-branes, such that~\cite{Dasgupta:1996yh,Sethi:1996es,Dasgupta:1999ss}
\begin{equation}
    \frac{\chi(X_4)}{24}= n_{\rm D3} + \frac12 \int_{X_4} G_4\wedge G_4 \,,
\end{equation}
where $n_{\rm D3}$ is the number of spacetime-filling D3-branes and $\chi(X_4)$ the Euler characteristic of $X_4$.
Instead, for the local, non-compact models, there is no analogue of the Euler characteristic. Still, even in the non-compact models, we can consider the net change in D3-brane charge induced by the blow-up. To that end, we recall the expression for the Euler characteristic for compact Calabi--Yau fourfolds in terms of the Hodge numbers ($h^{1,1}, h^{2,1}, h^{3,1}$) of the fourfold~\cite{Klemm:1996ts}
\begin{equation}\label{chi}
\frac{\chi(X_4)}{24} = \frac14\left(8+h^{3,1}+ h^{1,1} - h^{2,1}\right)\,.
\end{equation} 
In a compact model, the LHS has to be canceled by fluxes or spacetime-filling D3-branes. For the non-compact models that we consider here, we do not necessarily have to cancel the tadpole.  Nevertheless, since the system features enhanced supersymmetry, locally the blow-up is not sensitive to global data and no additional D3-brane charge is induced locally.\footnote{Otherwise, the system would not have a decoupled $\cN=2$ sector, as $\cN=1$ physics from the bulk would enter the field theory description and spoil the enhanced supersymmetry.} Let us introduce the following quantities capturing the additional deformations of $\widehat{X}_4$ 
\begin{equation}
 \delta h^{p,q} = h^{p,q}(\widehat{X}_4) - h^{p,q}(X_4)  \,. 
\end{equation} 
The difference in the Euler characteristic between $X_4$ and $\widehat{X}_4$ can then be expressed as 
\begin{equation}\label{deltachi}
 \delta \chi = \chi(\widehat{X}_4) - \chi(X_4) = 6\left(\delta h^{3,1}+\delta h^{1,1} - \delta h^{2,1} \right)\,.
\end{equation} 
As the birational factorization introduces an additional divisor on $\widehat{B}_3$, the smooth Weierstrass model over $\widehat{B}_3$ has an additional vertical divisor as well, such that $\delta h^{1,1}=1$. The contribution to $\delta \chi$ of the additional divisor can be canceled simply if $\delta h^{2,1}=1$. From a physical perspective, such a change in $h^{2,1}$ can be motivated based on the local supersymmetric enhancement as the following three arguments show:\footnote{As we discuss in section~\ref{ssec:D3moduli}, the situation is different if we blow up a point $p\in B_3$.}
\begin{enumerate}
    \item The blow-up introduces two additional $\cN=1$ chiral fields consisting of one saxionic direction (the volume of $E$) and three axion directions: $\Im\,T_E$ and two axions from $$\delta h^{2,1}=\delta h^{1,2}=1\,,$$ obtained by reducing the M-theory three-form over the associated three-cycles.\footnote{After the blow-up, the number of \textit{geometric} moduli increases by one, since F-theory has no geometric moduli associated with $h^{2,1}$.} As we explain in more detail below, these four additional scalar fields are the degrees of freedom required to form 4d $\cN=2$ supermultiplets. Therefore, due to the enhanced supersymmetry, we expect the blow-up to introduce an additional $h^{2,1}$.
    \item The local physics around $C_0$ and its blow-up cannot depend on the local D3-brane charge. The reason is that, locally, the geometry resembles a compactification of Type IIB string theory on a Calabi--Yau threefold, which cannot have any D3-brane charge. Similarly, the local geometry is insensitive to a change in $\delta h^{3,1}$. This is because the birational factorization discussed previously exists and in the non-compact case can be realized at any point within the complex structure moduli space since non-compactness of the Calabi–Yau manifold makes it possible to send global phenomena arbitrarily far from the local $C_0$ geometry.\footnote{Recall that, so far, we worked with a \emph{local} model for which the D3-brane tadpole does not have to be canceled and the Euler characteristic of the fourfold is not well-defined. As we will see in section~\ref{ssec:global}, if we embed this local model into a compact threefold, the Euler characteristic of the fourfold changes, leading to a significant decrease in the number of complex structure deformations $\delta h^{3,1} \ll 0$. However, since this effect arises only in the compact case, we will ignore it for our further analysis of the local, non-compact case.} We will further elaborate on this in section \ref{ssec:global}.
    \item The change in $\delta h^{2,1}=1$ for the transition $X_4\to \widehat{X}_4$ parallels the change in the Hodge number $h^{2,1}(X_3)$ of the threefolds describing the theories $\mathbb{T}_2$ and $\mathbb{T}_0$ discussed in section~\ref{sec:6d}. In six dimensions, the additional element in $H^{2,1}(X_3)$ provides a full hypermultiplet since there are geometric complex structure deformations associated with the corresponding 3-forms on $X_3$ in addition to the periods of $B_2$ and $C_2$ (or periods of $C_3$ in the dual M-theory compactification). In compactifications on fourfolds, the geometric deformations are absent, but the 2-form periods remain part of the massless spectrum of states.
\end{enumerate}

In summary, the local supersymmetry enhancement tells us that blowing up the curve $C_0$ with normal bundle as in~\eqref{normalbundleF} leads to $\delta h^{2,1}=1$. Compared to $X_4$, the fourfold $\widehat{X}_4$ therefore contains two additional three-cycles $\Gamma_1$ and $\Gamma_2$. Integrating the M-theory three-form $C_3$ over these cycles gives rise to the axions \cite{Denef:2008wq, Grimm:2010ks}
\begin{equation}
    b_1^E= \int_{\Gamma_1} C_3 \,,\qquad b_2^E = \int _{\Gamma_2} C_3\,,\label{chiralsh21}
\end{equation}
see also~\cite{Greiner:2015mdm,Greiner:2017ery} for a detailed discussion of three-form periods in F-theory. In the F-theory lift, these axions correspond to periods of $B_2$ and $C_2$ and are thus of a similar origin as the fields $\tilde b_2, \tilde c_2$ defined in~\eqref{TypeIIBfields}. As discussed above, since $\tilde{C}_0\in H_2^+(X_3)$, at the perturbative level these fields are projected out by the orientifold. Reintroducing these states as massless fields through the blow-up sounds contradictory at first sight. However, even though the existence of these fields can be inferred from the classical geometry of $\widehat{X}_4$, they do not have to be massless at the quantum level. 

In fact, both fields $b_{1,2}^E$ have a mass of order of the 10d Planck scale along the locus $\cM_{B_3}\subset \cM_{\hat B_3}$. To see that, we notice that the exceptional divisor $E\subset \widehat{B}_3$ is rigid, such that a Euclidean D3-brane instanton wrapped on $E$  can in principle contribute to the non-perturbative superpotential as 
\begin{equation}\label{eq:Wnp}
    W_{\rm n.p.} \supset \cA_E\, e^{-2\pi T_E}\,,
\end{equation}
where $\cA_E$ is the one-loop Pfaffian of the instanton on $E$. The Pfaffian can have a non-trivial dependence on all chiral multiplets apart from the K\"ahler moduli. In particular, as argued in~\cite{Witten:1996hc} for M-theory and~\cite{Ganor:1996pe} for F-theory, the Pfaffian non-trivially depends on the three-form axions in M-theory/two-form axions in F-theory. To determine the exact dependence of $\cA_E$ on the axions $b_{1,2}^E$, one would have to compute the five-brane partition function for the M5-brane instanton on $\pi^*(E)$ or use modular symmetry arguments as in~\cite{Grimm:2007xm,Chen:2025rkb} to fix the axion-dependence of the non-perturbative superpotential. We are not going to perform this calculation here, but notice that the zeros of $\cA_E=\cA_E(b_1^E,b_2^E)$ correspond to supersymmetric vacua of the theory. 

The non-trivial dependence of the $\cA_E$ on the axions effectively induces a non-zero mass for the fields $b_{1,2}^E$ set by the instanton action, i.e., 
\begin{equation}\label{massaxion}
m^2_{b_{1,2}^E} = \mathcal{O}\left(e^{-2\pi T_E} \right)M_{10}^2 \,.
\end{equation}
Along the locus $\{T_E=0\}\subset \cM_{\widehat{B}_3}$, the fields $b_{1,2}^E$ thus have a mass of order of the 10d Planck scale. This is consistent with the perturbative Type IIB orientifold picture since the orientifold action effectively gives a Planck-scale mass to the fields $(\tilde{b}_2,\tilde{c}_2)$, which is equivalent to projecting them out from the low-energy effective action. Consistency with the perturbative Type IIB picture further tells us that the supersymmetric vacuum of the non-perturbative superpotential in~\eqref{eq:Wnp} has to correspond to $b_1^E=b_2^E=0$ such that 
\begin{equation}
    W_{\rm n.p.} = 0 \qquad \text{for}\qquad b_1^E=b_2^E=0\,.
\end{equation}
Thus, also in the theory on $\widehat{B}_3$, the fields $b_1^E$ and $b_2^E$ are not dynamical but their vev is fixed to zero as in the Type IIB orientifold.\footnote{Notice that in the geometric regime ($\text{Re}\,T\to \infty$) of the K\"ahler moduli space of $\widehat{B}_3$ the two axions become approximately massless since the $e^{-2\pi T_E}$ factor in $W_{\rm n.p.}$ suppresses the overall mass scale. }

In contrast, the field $T_E$ remains a flat direction and hence a modulus of the theory. We can therefore give a vev to $T_E$ and move from the model based on $B_3$ to a model based on $\widehat{B}_3$. Together with the chiral multiplet associated with $t_0$ already present in the model based on $B_3$, the chiral multiplet $T_E$ provides the field content of a massless $\cN=2$ hypermultiplet in four dimensions. \newline 

To summarize, the model based on $B_3$ containing the curve $C_0$ with normal bundle as in~\eqref{normalbundleF} can be connected to a different geometry $\widehat{B}_3$ by performing the birational factorization of the flop for $C_0$. The model $B_3$ is then realized at a sublocus $\cM_{B_3}$ in the K\"ahler moduli space $\cM_{\widehat{B}_3}$ of $\widehat{B}_3$. The volume of the exceptional divisor associated with the blow-up $B_3\to \widehat{B}_3$ is part of an additional chiral multiplet that, together with the chiral multiplet containing the volume of $C_0$, forms the field content of a massless $\cN=2$ hypermultiplet. In order for the blow-up not to introduce D3-brane charge in the non-compact model, $\widehat{X}_4$ contains a three-form for which the periods of the M-theory three-form $C_3$ (or equivalently the Type IIB two-forms $C_2$ and $B_2$) yield an additional chiral multiplet. The vev of the scalar fields in this multiplet is set to zero along $\cM_{B_3}$ but becomes a massless scalar field in the large volume regime ($\text{Re}\,T_E\gg 0$) of $\widehat{B}_3$ as is clear from~\eqref{massaxion}.
\subsubsection{Phase structure of $\mathcal{M}_{\widehat{X}_4}$}
\begin{figure}[t!]
\hspace{-2em}
    \centering
    \includegraphics[width=0.7\linewidth]{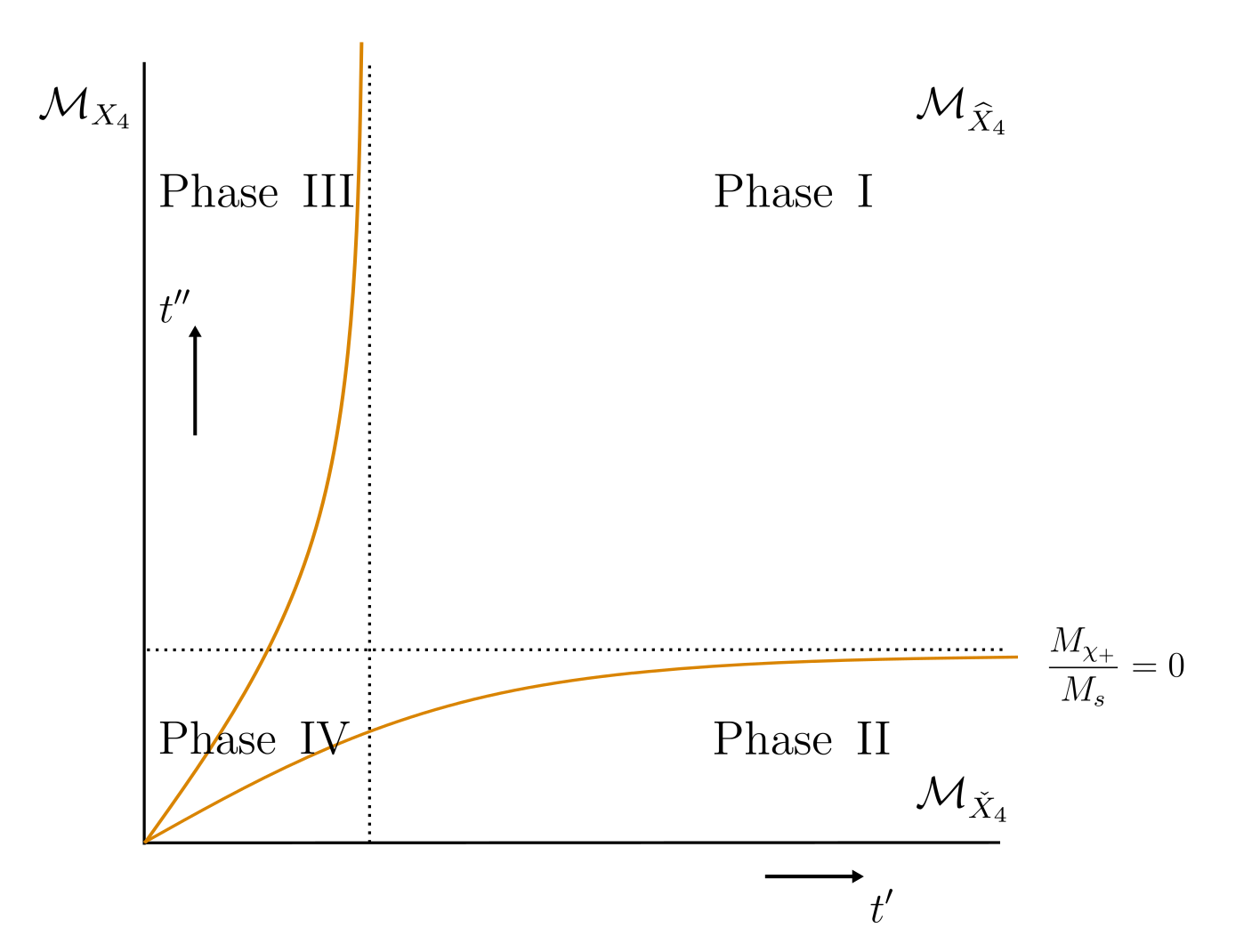}
    \caption{A sketch of the phase structure of $\mathcal{M}_{\widehat{X}_4}$. The Phase  \hyperref[phaseI]{I} corresponds to the classical geometric description of F-theory on $\widehat{X}_4$. Phases \hyperref[phaseII,III]{II and III} correspond to a regime in which the volumes of the cycles $C'$ and $C''$ differ hierarchically. The deep interiror of these phases can respectively be identified with the moduli space of the theory compactified on $\cM_{X_4}$ and $\cM_{\check{X}_4}$. Phase \hyperref[phaseIV]{IV} corresponds to a small-volume regime for both curves and marks the transition into the full non-perturbative regime. The orange line represents the locus along which the mass of $\chi_+$ (or $\check\chi_+$) vanishes, as discussed in Section \ref{sec: phase iv}.}
    \label{fig:phases}
\end{figure}

As is clear from the above discussion, the non-perturbative physics of the original theory defined on $X_4$ is captured by the theory on $\widehat{X}_4$. To extract the physics of the theory associated with $\widehat{X}_4$, it is useful to distinguish the different phases of its field space $\mathcal{M}_{\widehat{X}_4}$. The different phases correspond to different values for the K\"ahler moduli of $\widehat{X}_4$ which are given by $(t_0, c_4^{(0)})$ already present in $X_4$ and the complex scalar $(T_E)$ gained after the blow-up. The phase structure of $\cM_{\widehat{X}_4}$ is better studied in terms of the scalar fields arising directly from $C'$ and $C''$ defined as 
\begin{equation}\begin{aligned}
    t'&= \int_{C'} J_{\widehat{B}_3} \,, \quad t'' =  \int_{C''} J_{\widehat{B}_3}\,,\\
    a'&= \left(\int_{C'} C_4\right)^\vee \,,\qquad a''= \left(\int_{C''} C_4\right)^\vee\,. 
\end{aligned}
\end{equation} 
In terms of these scalars, we can distinguish four different phases as  (I) $t',t''\gg1$, (II) $t'\gg1\gg t''>0$, (III) $t''\gg1\gg t'>0$, and (IV) $1\gg t',t''> 0$, see Figure \ref{fig:phases}. \newline

\noindent\textbf{Phase I:}\label{phaseI} This phase corresponds to the regime $t', t''\gg1$. In this phase, the classical geometric description of the 4d $\cN=1$ theory in terms of F-theory on $\widehat{X}_4$ is valid. There is no local supersymmetry enhancement in Phase I, since the curves $C'$ and $C''$ intersect the anticanonical class $K_{\widehat{B}_3}$. Wrapping a D3-brane on either of these curves yields strings for which the worldsheet theory does not feature enhanced supersymmetry. The light degrees of freedom are given by two massless chiral multiplets and a massive $\cN=1$ vector multiplet. In addition, the fields $b_{1,2}^E$ are part of a chiral multiplet that is massive but with a mass of the order of $e^{-2\pi T_E}\ll 1$ which is thus exponentially small in this phase. \newline

\noindent\textbf{Phase II and III:}\label{phaseII,III} These phases represent the regimes $t'\gg1\gg t''>0$ (Phase II) and $t''\gg1\gg t'>0$ (Phase III), respectively. The chiral multiplet $(b_1^E, b_2^E)$ from \eqref{chiralsh21} has a mass of the order of the 10d Planck scale due to the non-perturbative superpotential generated by the Euclidean D3-brane instanton on $E$ such that their vacuum expectation values are fixed to zero. Accordingly, the light degrees of freedom are given by the two massless chiral multiplets corresponding to the K\"ahler moduli and the massive vector multiplet. The two massless chiral multiplets together form the degrees of freedom of a massless $\cN=2$ hypermultiplet such that the enhanced supersymmetry is indeed manifest in the massless spectrum. \newline

\noindent\textbf{Phase IV:}\label{phaseIV} Phase IV is the regime $1\gg t',t''> 0$, i.e., the small volume phase for both curves $C'$ and $C''$. For the original theory on $X_4$, Phase IV corresponds to the small volume regime for the curve $C_0$, i.e., $t_0\simeq 0$. In this regime, the subsector of the theory associated with $C_0$ is entirely decoupled from the bulk. Since this subsector has enhanced supersymmetry, the light spectrum associated with it has to fill out complete $\cN=2$ multiplets. Phase IV corresponds to a fully quantum regime of the theory, in which quantum effects originating from Euclidean D-branes wrapping the exceptional divisor $E$ as well as the two curves $C'$ and $C''$ become important. This phase will be the topic of the following subsection \ref{sec: phase iv}, where we will describe the light degrees of freedom make up a massless $\cN=2$ hypermultiplet and a massive $\cN=2$ vector multiplet that arise from a combination of perturbative and non-perturbative states.
\subsection{Light Degrees of Freedom in Phase IV}\label{sec: phase iv}
So far, we have focused on the transition between the model based on $X_4$ to the model based on $\widehat{X}_4$, concluding that there is an additional geometric massless field that augments the moduli space $\cM_{B_3}$ of the original model through the inclusion of (at least) one additional chiral multiplet. We now focus on the small volume regime for the curve $C_0$ in the original moduli space $\cM_{B_3}$ corresponding to Phase~\hyperref[phaseIV]{IV}. Recall that before orientifolding, the point $t_0=0$ corresponds to the origin of a field-theoretic Higgs branch for a $U(1)$ gauge theory. Specifically, this means that at $t_0=0$ a massive $\cN=2$ vector multiplet becomes massless. From the discussion in section~\ref{ssec:additionaldeformation}, we recall that in terms of $\cN=1$ multiplets, a massive $\cN=2$ vector splits as 
\begin{equation}
    (5,3)^{m\neq 0}_{\cN=2, {\rm V}} = (1,3)^{m\neq 0}_{\cN=1,{\rm V}} \,+ 2\times (2,0)^{m\neq 0}_{\cN=1, \rm C}\,.  \label{dof vector N=2}
\end{equation}
The explicit breaking of supersymmetry through the orientifold action projects out half of the degrees of freedom of the original $\cN=2$ multiplet. Here, we assume that the surviving $\cN=1$ multiplet is a massive vector multiplet which we denote by $\Upsilon$. 

As discussed above, in the small volume limit for $C_0$, the subsector of the theory featuring enhanced supersymmetry is entirely decoupled from the bulk. As a consequence of the decoupling and the local supersymmetry enhancement, the light degrees of freedom characterizing the decoupled subsector of the theory have to fill out full $\cN=2$ multiplets. In the following, we discuss how this is realized in Phase~\hyperref[phaseIV]{IV} of the theory. 

\subsubsection{Missing, massive degrees of freedom}
Apart from the massless $\cN=2$ hypermultiplet, we have to identify the light degrees of freedom in Phase~\hyperref[phaseIV]{IV} that together make up an $\cN=2$ vector multiplet that is massive at generic points in the moduli space. To achieve this, we need two chiral multiplets that, together with the massive $\cN=1$ multiplet $\Upsilon$, form the field content of the $\cN=2$ massive vector multiplet. We denote these two chiral multiplets by $\chi_0$ and $\chi_+$, reflecting that they are neutral and charged under $\Upsilon$, respectively. We thus have 
\begin{equation}\label{massivevectorsplit}
    (5,3)_{\mathcal{N}=2,\rm V}^{m\neq 0} = \Upsilon \oplus \chi_0 \oplus \chi_+\,. 
\end{equation}
These states cannot be of perturbative Type IIB origin since the orientifold action removes the corresponding states from the spectrum. For this reason, we must reconsider the non-perturbative completion of the theory, given by the model based on $\widehat{X}_4$. Working in the theory corresponding to $\widehat{X}_4$, we first have to understand the fate of the massive $\cN=1$ vector multiplet $\Upsilon$ as we transition from $X_4\to \widehat{X}_4$.

Recall that in the theory corresponding to Type IIB on the Calabi--Yau threefold $X_3$, the massive vector multiplet arises from reducing the Type IIB four-form $C_4$ over the three-chain $\Sigma_3$ with boundary containing $\tilde{C}_0$. This three-chain is also present after orientifolding as well as in the F-theory model based on the compactification on $X_4$. Accordingly, in the theory associated with $X_4$, the massive vector multiplet also arises from reducing $C_4$ over a three-chain $\Sigma_3$.  However, by blowing up the curve $C_0$ into the exceptional divisor $E$ the three-chain $\Sigma_3$ ceases to exist such that the massive vector multiplet cannot any longer arise from this three-chain. However, in the $\cN=1$ theory, the Higgsed $U(1)$ gauge field cannot disappear by a smooth deformation. Instead, it is just realized differently. 

To understand the realization of the Higgsed gauge field after the transition, we recall that upon performing the birational factorization of the flop, the first Chern class of the base threefold changes as in \eqref{shiftKbar}. Therefore, the exceptional divisor $E$ is contained in the canonical divisor of $\widehat{B}_3$, meaning that the O7-brane locus is sensitive to the divisor $E$. In fact, there is a component of the O7-brane locus wrapping $E$. Since the Weierstrass model over $\widehat{B}_3$ is smooth, there is no massless gauge theory associated with $E$. Instead, the would-be gauge theory arising from the 7-branes on $E$ is St\"uckelberg massive. The vector multiplet that arises from the 7-brane worldvolume theory is massive, and we can identify it with the massive $\cN=1$ vector multiplet $\Upsilon$. In other words, when transitioning from $B_3\to \widehat{B}_3$, a closed string massive $U(1)$ gets replaced by a massive open string $U(1)$. This is reminiscent of the discussion below \eqref{dof vector}, where the various states have different microscopic origins depending on whether the Higgs or the Coulomb branch perspective is considered. Notice that a similar replacement of a D-brane gauge theory by a closed string gauge theory has been discussed previously in the context of Type IIA/M-theory compactifications in~\cite{Atiyah:2000zz}.\footnote{Another example of a gauge theory having a perturbative closed string origin before and a brane origin after a smooth transition has recently been discussed in 4d $\cN=2$ compactifications of the heterotic string in~\cite{Monnee:2025msf}.}

We thus conclude that the origin of the massive vector boson differs before and after the blow-up. Before the transition, the massive mode arises from the reduction of the RR four-form $C_4$ over a three-chain. After the blow-up, however, it corresponds to a gauge boson living on the world volume of a 7-brane. At first glance, these appear to be fundamentally different origins.
However, within the framework of F-theory, this distinction is less sharp.   
This conclusion is further supported by considering the transition from F-theory to the Type IIB limit, or in other words, the transition $\widehat{X}_4\rightarrow X_4$. In that context, there are only two viable outcomes for an open string $U(1)$ in $\widehat{X}_4$:
\begin{itemize}
\item It could originate from a stack of D7-branes present before the blow-up, giving rise to an $SU(N) \times U(1)$ gauge theory. In this case, the $U(1)$ becomes Higgsed in the F-theory description after the blow-up.
\item Alternatively, it could descend from the closed string sector via the expansion of the RR four-form $C_4$ into its harmonic and non-harmonic components.
\end{itemize}
The first possibility is excluded by construction as a curve $C_0$ satisfying $\bar{K}_{B_3}\cdot C_0 = 0$ is considered. This ensures a locally constant axio-dilaton and thus precludes the presence of a D7-brane stack, as such configurations would induce a variation in the axio-dilaton. Therefore, the only possible realization of the Higgsed open string $U(1)$ after the blow-down is within the closed string sector.

\subsubsection{Charged States from D3-brane Strings}
The identification between closed and open string $U(1)$ already gives us an indication about the origin of the charged chiral multiplet $\chi_+$ that is needed to complete the field content of a massive $\cN=2$ vector multiplet. For a 7-brane gauge theory, the charged matter either arises at the intersection of 7-branes or from D3-branes wrapping curves that intersect or are contained in the 7-brane locus. 

Since charged modes arising at 7-brane (self-)intersections are typically of perturbative origin and massless, these cannot yield the mode $\chi_+$. Instead, $\chi_+$ has a non-zero mass, which is generically the case for the charged excitations of strings arising from D3-branes wrapping curves in $B_3$. In the original $\cN=2$ theory, the massive vector multiplet contains the degrees of freedom via an $\cN=2$ conifold transition are related to wrapped D3-brane states. From this perspective, it is not surprising that the charged chiral $\chi_+$ arises from D3-brane states.\footnote{Exchanging a D3-brane particle with an excitation of a D3-brane string parallels the replacement of a closed string (massive) $U(1)$ with an open string one.}

We thus have to identify a curve in $\widehat{B}_3$ that has a non-zero intersection with $E$ that can furthermore contribute a light state to the spectrum in the vicinity of $t_0=0$. A natural candidate is the curve $C''$ satisfying 
\begin{equation}
    E\cdot_{\widehat{B}_3} C''=-1 \,, 
\end{equation}
such that its excitations can carry charge under the broken $U(1)$ symmetry associated with $E$. Consider a D3-brane wrapping $C''$ leading to a string which we denote by $\mathtt{S}''$ (similarly, we denote the D3-brane on $C'$ by $\mathtt{S}'$). Classically, in Phase \hyperref[phaseI]{I}, the tension of this string is given by the volume of $C''$
\begin{equation}
    \frac{T_{\mathtt{S}''}}{M_{10}^2}\Bigg|_{\rm cl.} = \text{vol}(C'')M_{10}^2 \equiv t''\,.
\end{equation}
Thus, if we blow down $C''$, i.e. move deep into the interior of Phase~\hyperref[phaseII,III]{II}, the classical tension of the string vanishes. However, apart from the volume of $C''$, the tension of the string $\mathtt{S}''$ receives another contribution from the $B_2$-axion along $C''$ that has to be turned on since $\bar{K}_{\widehat{B}_3}\cdot C''\neq 0$~\cite{Freed:1999vc}
\begin{equation} 
 b_2'' \equiv \int_{C''} B_2 = \frac12 \,. 
\end{equation}
Therefore, the actual perturbative tension of the string is given by 
\begin{equation}\label{quantumtension}
\frac{T_{\mathtt{S}''}}{M_{10}^2}\Bigg|_{\rm pert.} = \left|t''+\frac{i}{2}\right|\,, 
\end{equation}
where the imaginary part accounts for the non-zero vev of $b_2''$. Notice that, due to this shift, the tension of the string is bounded as $T_{\mathtt{S}''}\geq \frac{1}{2}M_{10}^2 $ such that the string never becomes tensionless and, in fact, always has a tension of order of the 10d Planck scale. 

Still, the string can have finitely many excitations with mass below $M_{10}$. The reason is that due to the non-zero intersection with the anti-canonical bundle of $\widehat{B}_3$, the worldsheet theory of the string has a non-zero vacuum energy given by~\cite{DelZotto:2016pvm,Kim:2018gak,Lee:2018urn}
\begin{equation}
    E_0 = -\frac12 \bar{K}_{\widehat{B}_3}\cdot C''=-\frac12\,. 
\end{equation}
The mass of a string excitation is then given by\footnote{Here we assume that for low excitation levels $n$ the mass formula for the non-critical string ressembles that of a critical string.} 
\begin{equation}\label{massex}
    \frac{M_n^2}{M_{10}^2} = \left|\alpha \left(\frac{n}{\sqrt{2}} \frac{T_{\mathtt{S}''}}{M_{10}^2}+E_0\right)\right|^2 =\left|\alpha \left(\frac{n}{\sqrt{2}} \left((t'')^2 + \frac{1}{4}\right)^{1/2}-\frac{1}{2}\right)\right|^2\,. 
\end{equation}
We observe that the mass of the state with $n=1$ vanishes for $t''=1/2$, but remains non-zero for $n>1$. This means that along the locus $\{t''=\frac{1}{2}\}$ there is a massless state arising from the D3-brane string $\mathtt{S}''$ which we identify as the state $\chi_+$. Due to the symmetry between $C'$ and $C''$ the same analysis can be repeated for the string $\mathtt{S}'$. We denote the state for $n=1$ arising from $\mathtt{S}'$ by $\check{\chi}_+$. 

The expression for the tension of the D3-brane wrapped on $C''$ in \eqref{quantumtension} is valid in phase \hyperref[phaseII,III]{II} as it assumes $\text{vol}(C') M_{10}^2\gg 1$. In particular, this means that any non-perturbative effects associated with the curve $C'$ can be ignored. Away from this limit, the expression \eqref{quantumtension} can receive corrections of the form
\begin{equation}
    \frac{T_{\mathtt{S}''}}{M_{10}^2}\Bigg|_{\rm non-pert.} = \left|t''+ \beta \, e^{-2\pi t'}+\frac{i}{2}\right|\,,
\end{equation}
where the correction arises from $[p,q]$-string instantons wrapping $C'$. While we do not compute these corrections from first principle here, we notice that the requirement of having the d.o.f. making up a light $\cN=2$ vector multiplet in phase~\hyperref[phaseIV]{IV} implies $\beta \neq 0$. More precisely, we have to impose that the field $\chi_+$ becomes massless at the locus $\{t_0=0\}\subset \cM_{B_3}$ for it to form part of a $\cN=2$ vector multiplet that becomes massless at this locus. 

The state $\chi_+$ corresponds to the excitation level $n=1$. In terms of $t',t''$, the locus $\{t_0=0\}\subset \cM_{B_3}$ corresponds to $t'=t''=0$.\footnote{Recall that the locus $\{t''=0\}\subset \cM_{\widehat{B}_3}$ corresponds to the $t_0>0$ region of $\cM_{B_3}$.} Imposing that the string excitation with $n=1$ becomes massless at this point fixes 
\begin{equation}
    \beta = \frac{1}{2}\,. 
\end{equation}
such that in the vicinity of $t'=0\in \cM_{B_3}$ we have 
\begin{equation}
    \frac{m_{\chi_+}^2}{M_{10}^2}\Bigg|_{t''=0,t'\ll 1} = \alpha^2 \left(\left|\frac{1 - 2\pi t' + i}{2\sqrt{2}}  \right|-\frac12\right)^2 \simeq \frac{\alpha^2\pi^2 \left(t'\right)^2}{4} \,.
\end{equation}
As expected for a state that pairs with the massive $\mathcal{N}=1$ multiplet $\Upsilon$ to form a massive $\mathcal{N}=2$ multiplet, the mass of $\chi_+$ vanishes polynomially as $t' \to 0$. Thus, for $0 < t' \ll 1$ and $t'' = 0$, we identify the $n=1$ excitation of $\mathtt{S}''$ with the charged state $\chi_+$ appearing in~\eqref{massivevectorsplit}. In Figure~\ref{fig:phases}, we depict the loci in the $(t', t'')$-plane where this charged chiral becomes massless. As shown, quantum corrections shift the massless point to the origin, $t' = t'' = 0$.

Instead, for $t'=0$ and $0<t''\ll 1$, the state $\check\chi_+$ corresponding to the state $n=1$ of the string $\mathtt{S}'$ takes over the role of the charged chiral multiplet in the massive $\cN=2$ vector multiplet. Notice that we are not overcounting, since at $t'=t''=0$ the massive $\cN=1$ vector multiplet becomes massless and splits as 
\begin{equation}
     (1,3)^{m\neq 0}_{\cN=1, {\rm V}} = (0,2)^{m= 0}_{\cN=1,{\rm V}} \,+  (2,0)^{m= 0}_{\cN=1, \rm C}\,. \label{dof vector n=1}
\end{equation}
Depending on whether we turn on $t'$ or $t''$ and thus move to $t_0<0$ or $t_0>0$ along $\cM_{B_3}$, the state that pairs up with the $(0,2)$ massless vector is, respectively, the chiral multiplet $\chi_+$ or $\check\chi_+$. 

The origin of the chiral multiplets $\chi_+$ and $\check\chi_+$ thus depends on the sign of $t_0$. For $t_0 > 0$, $\chi_+$ arises as a string excitation of $\mathtt{S}''$, while $\check\chi_+$ belongs to the massive $\mathcal{N}=1$ vector multiplet associated with the Higgsed open string $U(1)$ on a 7-brane—in this phase, it originates from the 7-brane open string sector. Conversely, for $t_0 < 0$, $\check\chi_+$ arises as a string excitation of $\mathtt{S}'$, and $\chi_+$ is instead part of the massive vector multiplet.\footnote{The different origin of the same state is the analogue of what happens for conifold transitions (as discussed below \ref{dof vector}), where the origin of the degrees of freedom changes when moving from the Higgs to the Coulomb branch and vice versa.} The above can be summarized in the following split of the degrees of freedom comprising the massive 4d $\cN=2$ vector multiplet: 
\begin{align}
    t_0>0:\qquad (5,3)_{\cN=2,\rm V}^{m\neq 0} &\to \Upsilon  \oplus \chi_+ \oplus \chi_0  = \underbrace{(\Upsilon_0\oplus \check\chi_+)}_{\rm 7-brane } \oplus \underbrace{\chi_+}_{\mathtt{S}''\;\text{excitation}} \oplus\; \chi_0 \\ 
     t_0<0:\qquad (5,3)_{\cN=2,\rm V}^{m\neq 0} &\to \Upsilon  \oplus \check\chi_+ \oplus \chi_0  = \underbrace{(\Upsilon_0\oplus \chi_+)}_{\rm 7-brane } \oplus \underbrace{\check\chi_+}_{\mathtt{S}'\;\text{excitation}} \oplus \;\chi_0\,, 
\end{align}
where, as before, $\Upsilon_0$ contains the degrees of freedom of a massless $\cN=1$ vector multiplet.

\subsubsection{Completing the local $\cN=2$ spectrum}
Finally, the uncharged, light chiral multiplet $\chi_0$ in phase \hyperref[phaseIV]{IV} can simply be identified with the multiplet containing the axions $b_1^E$ and $b_2^E$ as scalar fields. The reason is that, as we stressed above, in the vicinity of $t_0=0$ the subsector of the theory associated with the curve $C_0$ is entirely decoupled from the bulk theory and we effectively consider a compactification of Type IIB string theory on a local Calabi--Yau threefold. Such a theory has to contain two light axion fields associated with the periods of the Type IIB two-forms. As these axions are uncharged under any 7-brane gauge theory, they indeed yield a neutral chiral multiplet $\chi_0$. 

We argued above that in the large volume regime for the model based on $B_3$, i.e., in the phases \hyperref[phaseII,III]{II and III} the chiral multiplet associated with $b_{1,2}^E$ has a mass of order of the 10d Planck scale due to the non-perturbative superpotential induced by the Euclidean D3-brane on the exceptional divisor $E$, see~\eqref{massaxion}. The mass of this state gets, however, corrected by perturbative and non-perturbative effects for finite $t'$. By a similar reasoning as before, the non-perturbative effects arise from (non-BPS) $[p,q]$-string instantons on the curves $C'$ and $C''$. From the low-energy effective action, these effects on the mass can be interpreted as non-perturbative corrections to the K\"ahler potential that give an additional contribution to the mass of the scalars $b_{1,2}^E$ that, at $t_0=0$, has to cancel the contribution already present in the phases \hyperref[phaseII,III]{II and III}.
Schematically, the mass of $\chi_0$ thus has the form 
\begin{equation}\label{mchi0}
    \frac{m_{\chi_0}^2}{M_{10}^2} \simeq  \alpha e^{-2\pi T_E}\left(1 - e^{-2\pi (t'+t'')}\right)^2\,,
\end{equation}
for some constant $\alpha$. The mass of the $\chi_0$ multiplet thus vanishes polynomially for $t',t''\to 0$, i.e., as we approach $t_0=0$ in the deep interior of phase \hyperref[phaseIV]{IV}. Let us stress that, again, we did not derive~\eqref{mchi0} from first principles. Instead, we were guided by the physics that arises at the origin of the Higgs branch of a 4d $\cN=2$ field theory. Imposing the existence of a massless, uncharged $\cN=1$ chiral multiplet then led to the expression~\eqref{mchi0}.

Using a combination of field theory and string theory input, we thus identified all light states in the vicinity of $t_0=0\in \cM_{B_3}$, or equivalently, $\{t'=t''=0\}\in\cM_{\hat B_3}$. On the one hand, we have the two classically massless chiral multiplets with scalar components $T_E$ and $t_0$ that provide the field content of the massless $\cN=2$ hypermultiplet. On the other hand, we have the excitations of the D3-brane on $C''$ (or $C'$) and the chiral multiplet associated with the periods of $B_2$ and $C_2$ that, together with the massive $\cN=1$ vector $\Upsilon$ arising from the massive gauge boson on the 7-brane, form the field content of an $\cN=2$ massive vector multiplet. 

Notice that also in the $\cN=2$ theory obtained as Type IIB string theory on $X_3$, the light degrees of freedom in the vicinity of $t_0=0$ arise from the geometric moduli, periods of the Type IIB two-forms, and, depending on the perspective, D3-brane states. Thus, the states that we identified in the analysis above are in no way exotic since we also just had to consider geometric deformations of $B_3\to \widehat{B}_3$, periods of the two-forms, and massive states arising from wrapped D3-branes. However, they appeared in slightly different ways in the light spectrum.

\subsection{Global Effects}\label{ssec:global}
So far we have considered the sector of the theory associated with $C_0$ as a local model decoupled from gravity. We are now interested in the imprints on the local physics from coupling this subsector back to gravity. This coupling to gravity is achieved by embedding $C_0$ in a compact Calabi--Yau fourfold. There are two effects of this global embedding to consider:
\begin{enumerate}
    \item  By going from a non-compact to a compact model, the D3-brane tadpole has to be canceled. In particular, the Euler characteristic of a compact Calabi--Yau fourfold changes under the blow-up of the curve $C_0$ into the exceptional divisor $E$. The change in the Euler characteristic can be associated with global effects of the compactification.
    \item The global breaking of $\cN=2\to \cN=1$ affects the local model.  In non-compact models, the effects that break $\cN=2\to \cN=1$ can be pushed infinitely far away from the location of the curve $C_0$ and can therefore be ignored. However, this is not possible in compact models where gravity mediates the breaking $\cN=2\to \cN=1$.
\end{enumerate}

In Section~\ref {sssec:tadpole}, we first discuss the topological consequences of global models, whereas in Section~\ref {sssec:SUSYbreaking}, we analyze the effect of global supersymmetry breaking on the states in subsectors with enhanced supersymmetry. 
\subsubsection{Tadpole Cancellation}\label{sssec:tadpole}

Instead of a non-compact model, we now consider a \emph{compact} base $B_3$ containing a flop curve $C_0$ as discussed above.\footnote{By a slight abuse of notation, we continue to refer to the bases as $B_3$ and $\widehat{B}_3$ and the smooth Weierstrass models over them as $X_4$ and $\widehat{X}_4$ which are now all assumed to be compact.} Again, we blow up the curve $C_0$ into an exceptional divisor $E$ and investigate the properties of the resulting theory. A significant difference from the non-compact case discussed above is that for the compact model, the D3-brane tadpole must be canceled. To see the effect of the tadpole cancellation condition on the physics of the blow-up, we first compute the change in the Euler characteristic of the fourfold. Assuming that both $X_4$ and $\widehat{X}_4$ are smooth Weierstrass models over $B_3$ and $\widehat{B}_3$, respectively, we have~\cite{Sethi:1996es}
\begin{equation}\label{eq:chifrombase}
    \chi(X_4) = 12 \int_{B_3} \left(c_1(B_3) c_2(B_3) +30 c_1(B_3)^3\right)\,,
\end{equation}
with an analogous expression for $\chi(\widehat{X}_4)$. Blowing up the curve $C_0$ introduces the exceptional divisor $E = \mathbb{P}(\cN_{C_0/B_3})$ which has the topology of $\mathbb{F}_0$. Denoting this blow-up by $\phi:\widehat{B}_3 \to B_3$, the Chern classes of $\widehat{B}_3$ and $B_3$ are related via
\begin{equation}
   c_1(\widehat{B}_3) =  \phi^* c_1(B_3) -E\,,\quad c_2(\hat B_3) = \phi^*c_2(B_3) - \phi^*(c_1(B_3)) \cdot E +[C_0]\,. 
\end{equation}
To compute the change in the Euler characteristic, we use 
\begin{equation}\begin{aligned}
    \phi^*\alpha\cdot  E &= 0 \,,\quad \phi^*\beta  \cdot E^2 = - \int_{C_0} \beta|_{C_0}\,,\\ E^3 &= - \int_{C_0} c_1(\cN_{C_0/B_3}) = -\text{deg}(\cN_{C_0/B_3}) = E^2 \cdot [C_0]\,. 
\end{aligned}\end{equation}
We then find
\begin{equation}\begin{aligned}
    \int_{\hat B_3} c_1(\hat B_3) c_2(\hat B_3) =  \int_{\hat B_3}&\Big[\phi^*c_1(B_3) \phi^* c_2(B_3) - E \phi^*c_2(B_3) - E (\phi^*c_1(B_3))^2  \\& + E^2 \phi^*c_1(B_3) + \phi^*c_1(B_3) [C_0] -E[C_0]\Big]\\
     = \int_{B_3}& c_1(B_3) c_2(B_3) -2 \int_{C_0} c_1(B_3) +  \int_{C_0}c_1(\cN_{C_0/B_3}) \,,
\end{aligned}\end{equation}
which for the curve with $C_0\cdot \bar{K}_{B_3}=0$ and $\cN_{C_0/B_3}=O(-1)\oplus O(-1)$ evaluates to 
\begin{equation}
    \int_{\hat B_3} c_1(\hat B_3) c_2(\hat B_3) = \int_{B_3}c_1(B_3) c_2(B_3) -2 \,. 
\end{equation}
Similarly, we evaluate
\begin{equation}\begin{aligned}
   \int_{\hat B_3} c_1(\hat B_3)^3 &= \int_{B_3} c_1(B_3)^3 + 3 \int_{\hat B_3} c_1(B_3) E^2  - \int_{\hat B_3} E^3 \\ 
   &= \int_{B_3} c_1(B_3)^3 - 3 \int_{C_0} c_1(B_3) + \deg(\cN_{C_0/B_3}) \,,
\end{aligned}
\end{equation}
which for a curve $C_0$ with normal bundle $\cO(-1)\oplus \cO(-1)$ yields 
\begin{equation}
    \int_{\hat B_3}c_1(\hat B_3)^3 = \int_{B_3} c_1(B_3) -2 \,. 
\end{equation}
Using \eqref{eq:chifrombase} we then find 
\begin{equation}
    \delta \chi = 12 \left(-2 - 2\times 30\right) = -744\,.
\end{equation}
Thus, as a consequence of the blow-up of $C_0$, the Euler characteristic decreases, implying that the number of spacetime-filling D3-branes also decreases\footnote{In this section, we assume that the entire D3-brane tadpole is cancelled by spacetime-filling D3-branes.}
\begin{equation}\label{eq:deltand3}
    \delta n_{D3} = -\frac{744}{24} =-31 \,. 
\end{equation}
Using the notation in~\eqref{deltachi}, this amounts to 
\begin{equation}
    \delta h^{1,1} +\delta h^{3,1} -\delta h^{2,1} = -124\,.
\end{equation}
In our analysis of the local model, we argued that the blow-up induces the changes 
\begin{equation}
    \delta h^{1,1} = \delta h^{2,1} =1 \,,
\end{equation}
such that the change in the Euler characteristic is accounted for by 
\begin{equation}\label{eq:deltah31}
    \delta h^{3,1} = -124\,. 
\end{equation}
Thus, due to the blow-up of the curve $C_0$, the fourfold $\widehat{X}_4$ has 124 complex structure deformations fewer than $X_4$. At first sight, this indicates that the blow-up is only possible at special loci in the complex structure moduli space of $X_4$ thus making the transition $X_4\to \widehat{X}_4$ reminiscent of a conifold transition in Calabi--Yau threefolds. In the following, we will argue that this is not the case. The key insight for this is that, unlike for F-theory in six dimensions or Type II compactifications on Calabi--Yau threefolds, the physical theory is not entirely determined by the geometry but requires additional ingredients such as spacetime-filling D3-branes. 

Let us focus on the dual M-theory description of our 4d F-theory compactification corresponding to the 3d $\cN=2$ theory obtained from M-theory on $X_4$. In this description, the role of the D3-branes is played by spacetime-filling M2-branes. The theory on a single M2-brane is a 3d $\cN=2$ field theory decoupled from gravity with a scalar field space of dimension eight given by $X_4$. In the theory on $X_4$, we can now consider the 31 M2-branes that have to be removed in the transition $X_4\to \widehat{X}_4$. Their combined moduli space has real dimension
\begin{equation}\label{dimM2}
\text{dim}_{\mathbb{R}}(\cM_{\rm M2}) = 8 \times 31 =  248 \,. 
\end{equation}
Apart from the M2-branes, certain degenerations inside the Calabi--Yau fourfold also signal the presence of field theory sectors. These arise, for example, if the Calabi--Yau fourfold is itself a non-trivial fibration of a Calabi--Yau threefold $Y_3$ over $\mathbb{P}^1$. Since $\mathbb{P}^1$ is compact, the threefold $Y_3$ has to undergo degenerations over points $p_1,\dots, p_n\in\mathbb{P}^1$ corresponding to infinite distance limits in the moduli space of $Y_3$. The points $p_1,\dots, p_n$ then host localized field theory sectors that are decoupled from gravity.\footnote{These are the analogues of e.g., little string theories arising in 6d realizations of F-theory at degenerations of the fiber of a K3-fibered Calabi--Yau threefold.} The relative positions of the $p_1, \dots, p_n$ correspond to complex structure deformations of $X_4$. Let us stress that in the above example, the presence of these degenerations is a consequence of the base $\mathbb{P}^1$ being compact. Otherwise, the fiber $Y_3$ would not necessarily have to undergo such degenerations at a finite location because the degeneration points could be sent to infinity, $p_i\to \infty$.  

Thus, a subset of the complex structure deformations of the fourfold is associated with a decoupled field theory sector that arises as a consequence of the compact topology of $X_4$.\footnote{This subsector is not to be confused with the decoupled subsector associated with the curve $C_0$. This latter subsector exhibits enhanced supersymmetry, unlike the one referred to here.} Together with the M2-branes, these form a field theory sector within the 3d $\cN=2$ theory of gravity, which can be viewed as the analogue of the set of SCFTs and LSTs arising in 6d compactifications of F-theory. When transitioning from $X_4\to \widehat{X}_4$, this entire field theory sector has to form an interacting field theory sector that decouples from the rest. 

Na\"ively, such an interacting field theory sector arises at a locus $\cM_{\rm int}\subset \cM_{\rm c.s.}(X_4)$ with codimension 
\begin{equation}
    \text{codim}_{\mathbb{R}}(\cM_{\rm int}\subset\cM_{\rm c.s.})=248\,. 
\end{equation}
This suggests that the field theory configuration has to be very special. However, in this case, the M2-branes can still be at generic positions such that the effective moduli space has again real dimension 248 as in~\eqref{dimM2}. Since the field theory sector associated with the 124 complex structure deformations is on the same footing as the field theory sector associated with the 31 M2-branes, the converse also has to be possible: If the 31 M2-branes are in a special configuration, the transition $X_4\to \widehat{X}_4$ occurs for \emph{any} value of the complex structure moduli. We can thus consider the field theory moduli space $\cM_{\rm FT}\subset \cM_{\rm c.s.} \times \cM_{\rm M2}\,$, which has codimension
\begin{equation}
    \text{codim}_{\mathbb{R}}(\cM_{\rm FT}\subset\cM_{\rm c.s.}\times \cM_{\rm M2})=248\,. 
\end{equation}
Unlike $\cM_{\rm int}$, the actual field theory moduli space $\cM_{\rm FT}$ can have codimension 0 inside $\cM_{\rm c.s.}$. Physically, the spacetime-filling M2-branes thus ensure that the transition is possible for any complex structure as expected on physical grounds. Let us stress that this is very different from analogue (extremal) transitions in theories with minimal supersymmetry in six dimensions, which are only possible for special choices of the complex structure (conifold points). The observation that the transitions in 4d $\cN=1$ theories discussed here are possible at any point in the complex structure moduli space thanks to the presence of spacetime-filling D3/M2-branes highlights that in genuine 4d $\cN=1$ theories all sectors of the moduli space (K\"ahler, complex structure, and D3-brane moduli) are not decoupled but in fact interact with each other.

To summarize, when performing the blow-up $X_4\to \widehat{X}_4$ for a compact fourfold, the M2-branes and complex structure deformations give rise to an interacting field theory sector macroscopically away from the curve $C_0$. The moduli associated with this field theory sector are lifted as we move to the geometric description of $\widehat{X}_4$. To consistently connect the physics associated with $X_4$ to the large volume regime of $\widehat{X}_4$, it is crucial that in F-/M-theory realizations of 4d $\cN=1$ theories/ 3d $\cN=2$ theories, the geometry of the compactification manifold does not entirely determine the physics. We already saw in the local description that the geometric phase of $\widehat{X}_4$ (Phase \hyperref[phaseI]{I}) does not contain the locus $\cM_{B_3}$, but the latter is realized in the deep interiror of the phases~\hyperref[phaseII,III]{II~and~III} of $\cM_{\widehat{B}_3}$. Thus, in the field space of $\widehat{X}_4$ the geometric regime of $\widehat{X}_4$ and the geometric regime of $X_4$ are separated by a finite distance corresponding to the diameter of the phases~\hyperref[phaseII,III]{II~and~III}. As we traverse these phases, the moduli associated with $\cM_{\rm FT}$ are lifted. The fact that phases~\hyperref[phaseII,III]{II~and~III} correspond to the deep quantum regime where the geometric description of the theory on $\widehat{X}_4$ is heavily corrected, reflects that also not all moduli in $\cM_{\rm FT}$ are of geometric origin.

\subsubsection{Global Supersymmetry Breaking}\label{sssec:SUSYbreaking}
When considering a compact base, effects breaking $\cN=2\to \cN=1$ in the full theory of quantum gravity will be mediated into the subsector with enhanced supersymmetry as finite-volume effects. The relevant control parameter in this context is the action of a Euclidean D3-brane instanton wrapping a non-exceptional, movable divisor $D_{\rm mov}$. More precisely, we define the parameter 
\begin{equation}
    T_{D,{\rm mov}} = \frac{1}{2}\int\limits_{D_{\rm mov}}J_{B_3}^2 \,,
\end{equation}
where $J_{B_3}$ is the K\"ahler form on $B_3$ measured in units of $M_{10}$. The mediation of the supersymmetry breaking effects has to be via non-perturbative effects arising from Euclidean D3-brane instantons wrapping $D_{\rm mov}$ that correct the K\"ahler potential. Before we turn to the effect of supersymmetry breaking $\cN=2\to \cN=1$, let us see how the K\"ahler potential mediates the effect of supersymmetry breaking on subsectors with enhanced supersymmetry in theories of gravity with $\cN=2$ supersymmetry.

\paragraph{Analogue with extended supersymmetry.} 
 Consider Type IIA string theory compactified on a Calabi--Yau threefold that is an elliptic fibration over $\mathbb{P}^2$. An example of this kind is the resolved degree-18 hypersurface in $\mathbb{P}_{1,1,1,6,9}$~\cite{Candelas:1994hw}. Here, the elliptic fiber can be viewed as a sector with enhanced supersymmetry. The reason is that locally in the vicinity of each fiber, the base $\mathbb{P}^2$ is flat such that we have enhanced $\cN=8$ supersymmetry. The supersymmetry is globally broken through the non-trivial fibration $\cE\to \mathbb{P}^2$. In the limit of large $\text{vol}(\mathbb{P}^2)$, the supersymmetry-breaking effects are diluted and for a given generic fibral curve can be pushed arbitrarily far away from this curve. Let us denote by 
\begin{equation}
    \tau_\cE = \int_{\mathcal{E}}(B_2+iJ) \,,
\end{equation}
the K\"ahler modulus corresponding to the complexified volume of $\cE$. The enhanced supersymmetry is reflected in the fact that the moduli space for $\tau_\cE$ is the upper half plane modded out by a subgroup of $SL(2,\mathbb{Z})$. As in the previous discussion, we are interested in the small volume regime for the curve exhibiting enhanced supersymmetry, corresponding here to $\tau_\cE\to 0$. This point is classically part of the moduli space. At this point, there are three kinds of states that have vanishing mass corresponding to the D2-brane on $\cE$, the D4-brane on the vertical divisor $\cE\to h$ with $h$ the hyperplane class of $\mathbb{P}^2$, and the D6-brane wrapping the entire threefold. Let us introduce the coordinate $q_\cE$ that is classically related to $\tau_\cE$ via the mirror map 
\begin{equation}
    j(\tau_\cE) = \frac{1}{1728q_\cE(1-432q_\cE)}\,,
\end{equation}
where the LHS is expressed in terms of the $SL(2,\mathbb{Z})$-invariant $j$-function. The above expression reflects that, classically, the $\tau_\cE$ moduli space is a double-cover of the $SL(2,\mathbb{Z})$ fundamental domains with cusps at $q_\cE=0$ and $q_\cE=1/432$. Let us denote the complexified K\"ahler modulus of the hyperplane class $h\subset \mathbb{P}^2$ by $\tau_h$. In the large $\tau_h\to i\infty$ limit, the principal component of the discriminant locus of $\mathbb{P}_{1,1,1,6,9}[18]$ is given by 
\begin{equation}
    \Delta_{P,0} = (1-432q_\cE)^3 = 0\,. 
\end{equation}
Thus, the cusp at $q_\cE=1/432$ corresponds to a triple root of $\Delta_P$ reflecting the fact that, classically, there are three kinds of states becoming massless at that point. The situation is different if we set $\tau_h$ to a finite value and thus do not entirely dilute the supersymmetry-breaking effects. Let us introduce the coordinate $q_h$ such that, for large $\text{Im}\,\tau_h$, 
\begin{equation}
    q_h = e^{2\pi i \tau_h}\,.
\end{equation}
For $q_h\neq0$, the expression for $\Delta_P$ receives corrections and takes the form~\cite{Morrison:1994fr,Cota:2019cjx}
\begin{equation}
    \Delta_P= (1-432q_\cE)^3-432^3\cdot 27q_\cE q_h =0 \,.
\end{equation}
The change in the expression for the principal discriminant can be associated to corrections in the K\"ahler potential for the moduli in the vector multiplets of the 4d $\cN=2$ effective action. The effect of $q_h\neq 0$ and hence finite supersymmetry breaking effects on the K\"ahler potential is that the triple root of $\Delta_{P,0}$ splits into three simple roots of $\Delta_P$. We specifically notice that the first singularity arises for finite values of $\text{Im}\,\tau_\cE$ which are exponentially suppressed in $\tau_h$~\cite{Wiesner:2022qys}. To see this, we recall that $q_\cE=1/432$ corresponds to $\tau_E=0$. Solving $\Delta_P=0$ for $q_\cE$ we then find
\begin{equation}
    1 -2\pi \text{Im}(\tau_\cE) = 1 -3 q_h^{1/3} + \mathcal{O}(q_h^{2/3})\,.  
\end{equation}
At $\Delta_P=0$, the central charge of the D6-brane wrapping the entire threefold vanishes. Thus, even though classically, the mass of the D6-brane and the D2-brane on $\cE$ (as measured by $|\tau_E|$) are proportional to each other, the supersymmetry-breaking effects induced by $q_h\neq0$ give an exponential mass hierarchy between the two states in the small volume regime for $\cE$. 

To summarize, the example of $\mathbb{P}_{1,1,1,6,9}[18]$ illustrates that supersymmetry breaking effects lead to exponentially suppressed corrections in subsectors with enhanced supersymmetry. Moreover, in this example there are three kinds of $\cN=2$ multiplets that, if we decouple the supersymmetry breaking effects entirely, become massless at the small volume point for the curve exhibiting enhanced supersymmetry. Instead, if the supersymmetry-breaking effects are coupled back in, i.e., for $\tau_h<\infty$, the three $\cN=2$ states do not become light at the same point. The difference in the mass of these states is exponentially small in $\tau_h$. Moreover, the D2-brane on $\cE$, that is classically the lightest in the small volume regime, does not become massless for finite $\tau_h$.  \newline 

Coming back to our $\cN=1$ setup, we realize that if we decouple the supersymmetry breaking effects, there are also three $\cN=1$ multiplets becoming massless at $t=0$: the massive $\cN=1$ vector multiplet $\Upsilon$ as well as the two chiral multiplets $\chi_+$ and $\chi_0$. Given the three D-brane states discussed in the above higher-supersymmetric Type IIA compactification, we view $\Upsilon$ as the analogue of the D2-brane on $\cE$, i.e., the state that we expect to classically become massless at the vanishing locus for $C_0$ . The charged chiral, $\chi_+$, is instead the analogue of the D6-brane on $X_3$ and $\chi_0$ the analogue of the D4-brane on the vertical divisor. 

For $T_{D,\rm mov}<\infty$, the supersymmetry breaking effects are non-vanishing. In the subsector with enhanced supersymmetry, the effect of finite $V_{D,\rm mov}$ is of the order $e^{-T_{D,\rm mov}}$ paralleling the effect of finite $\tau_h$ in the Type IIA discussion above. For finite $T_{D,\rm mov}$, the locus $\Delta_0=\{t_0=0\}$ at which three states become massless splits into three components. One of these three components corresponds to the locus 
\begin{equation}
    \Delta=\{t_0 = \gamma e^{-T_{{D,\rm mov}}}\}\subset \cM_{B_3}\,\quad \text{for}\quad 0<\gamma\sim \mathcal{O}(1)\,. 
\end{equation}
At this locus, the chiral multiplet $\chi_+$ is massless. Instead, the $\cN=1$ vector multiplet and the state $\chi_0$ remain massive with masses of order 
\begin{equation}\label{eq:massachi0}
    m_{\Upsilon,\chi_0}\Big|_{\Delta} \sim \mathcal{O}(e^{-T_{{D,\rm mov}}})\,. 
\end{equation}
Recall that the multiplets $(\Upsilon,\chi_+,\chi_0)$ together form the field content of a would-be massive $\cN=2$ vector multiplet. If $\cN=2$ supersymmetry was unbroken globally, the mass of all three $\cN=1$ multiplets would be the same. If, instead, we allow for finite $T_{D,{\rm mov}}$, the masses of these $\cN=1$ multiplets differ by an amount exponentially small in $V_{D,{\rm mov}}$. Thus, the enhanced supersymmetry is very mildly broken to minimal supersymmetry, as expected for gravitationally induced supersymmetry breaking. 

The uncharged chiral multiplet does not interact at all with the vector multiplet $\Upsilon$. Therefore, its mass is not affected by non-trivial dynamics of $\Upsilon$ if we turn on the supersymmetry breaking effects. The mass $m_{\chi_0}$ hence still vanishes at $t_0=0$ where $\Upsilon$ and $\chi_+$ have equal but finite mass exponentially small in $V_{D,\rm mov}$. As we move to $t_0<0$, the field $\chi_+$ arising from the string $\mathtt{S}_{C''}$ provides the longitudinal polarization of $\Upsilon$ as well as scalar in the massive vector multiplet. In contrast, the field $\check{\chi}_+$ splits off from the vector multiplet to give a chiral multiplet in its own right, which we interpreted as arising from an excitation of the string $\mathtt{S}_{C'}$. The last component of the split of $\Delta_0$ is given by 
\begin{equation}
   \check\Delta=\{t_0 = -\gamma e^{-T_{{D,\rm mov}}}\}\subset \cM_{B_3}\,\quad \text{for}\quad 0<\gamma\sim \mathcal{O}(1)\,,
\end{equation}
along which the chiral multiplet $\check\chi_+$ becomes massless. Along $\check\Delta $ the masses of $\Upsilon$ and $\chi_0$ are parametrically of the same order as in \eqref{eq:massachi0}. 

Let us emphasize that, just like the D2-brane wrapped on $\mathcal{E}$ in the Type IIA setup discussed above, the vector $\Upsilon$ is never massless as a consequence of the supersymmetry-breaking effects. Since the charged fields $\chi_+$ or $\check\chi_+$ are the longitudinal components of the massive $\mathcal{N}=1$ multiplet and one of them is always massive, it follows that the field $\Upsilon$ must also remain massive throughout the field space. Thus, for finite $T_{D,\rm mov}$ the supersymmetry-breaking effects removes the point at which an additional $U(1)$ gauge symmetry arises. 

Finally, we should emphasize that we have expressed $V_{D,\text{mov}}$ in the Einstein frame. Therefore, the large volume limit $V_{D,\text{mov}}$ can be taken by either going to weak string coupling or large volume in string units. Importantly, it is thus not sufficient to take the large volume limit if, at the same time, the string coupling is increased. An example of the interplay between these two limits can be observed in the instanton corrections arising from $(p,q)$-strings wrapping a curve $\mathcal{C}$, with an action,
\begin{equation}
    S_{p,q} \sim (\text{vol}(\mathcal{C}) M_{s}^2) |\tau q+p|\, ,
\end{equation}
where (unlike above) we measure the volume of the curve $\cC$ in string units. Whereas the instantons with $q=0$ decouple in the large volume limit for any value of the string coupling, suppression of the instanton effects with $q\neq 0$ also requires that the string coupling is sufficiently small. While the balance between overall volume and string coupling is important for quantifying supersymmetry breaking effects in concrete settings, we do not discuss this further here, as we are anyways only interested in the qualitative effects of gravity-induced $\cN=2\to\cN=1$ supersymmetry breaking.

Notice that apart from non-perturbative corrections to the mass of the states arising in the local sector with enhanced $\cN=2$ supersymmetry, there could be additional non-perturbative contributions to the superpotential. We already discussed the superpotential contribution of D3-brane instantons arising in the local model below equation~\eqref{eq:Wnp}. There we argued that in the local model with enhanced $\cN=2$ supersymmetry, the non-perturbative superpotential has to vanish identically. More generally, if we couple the local model back to gravity, there can be different field theory sectors that yield a non-zero contribution to the superpotential which further breaks supersymmetry $\cN=1\to \cN=0$. Without adding extra ingredients such as fluxes these lead to a runaway behavior of the scalar potential. The end-point of this runaway potential would then again be the large volume regime in which the supersymmetry breaking effects are diluted such that the analysis of the local system with enhanced $\cN=2$ supersymmetry remains unaffected.

\section{SUSY Enhancement in the Complex Structure Moduli Space}\label{sec:flux} 

In the previous section, we discussed local supersymmetry enhancement due to a geometric subsector featuring $SU(3)$ holonomy, which manifests in the curve $C_0$ not intersecting the anti-canonical divisor of the base $B_3$ in F-theory. In this section, we instead discuss the local appearance of eight supercharges through a local $SU(2) \times SU(2)$ holonomy.  To achieve this, we will focus on flux compactifications of F-theory on an elliptically fibered Calabi--Yau fourfold $X_4$. More specifically, we consider loci in the complex structure moduli space of $X_4$ that feature enhanced supersymmetry. The enhanced supersymmetry manifests in the periods of $X_4$ that along these loci can be expressed in terms of K3-periods, i.e. if a manifold with reduced $SU(2)$ holonomy group. This structure, featuring enhanced supersymmetry, enables the identification of configurations that become BPS along the locus and to understand their relevance in the context of domain wall transitions. \newline

In the heterotic/M-theory realization of the $\cN=(1,0)$ theory discussed in section~\ref{sec:6d}, we encountered two theories that differ by the $G_4$ flux quanta localized in the 9-branes at the end of the M-theory interval. The heterotic anomaly cancellation condition in this case translates to 
\begin{equation}
    Q^+_{G_4} + Q^-_{G_4}=0\,. 
\end{equation}
Viewed as a $G_4$ flux in an M-theory compactification on K3 to 7d, the $G_4$ flux breaks Poincar\'e invariance of the 7d theory as it leads to a non-trivial profile for the K3 along the M-theory interval. Reducing also along the interval leads to a 6d theory for which supersymmetry is preserved if the domain wall solution is BPS. The existence of such a BPS domain wall is also ensured by $Q^+_{G_4}=-Q^-_{G_4}$. 

In 4d F-theory compactifications on a Calabi--Yau fourfold $X_4$, we can also turn on $G_4$-flux, see \cite{Grana:2005jc,Denef:2007pq} for reviews on flux compactifications in string theory. The analogue of the heterotic anomaly cancellation is now the tadpole cancellation condition~\cite{Dasgupta:1996yh,Sethi:1996es,Dasgupta:1999ss}
\begin{equation}
    \frac{\chi(X_4)}{24}= n_{\rm D3} + \frac12 \int_{X_4} G_4\wedge G_4 \,,\label{eq: tadpole}
\end{equation}
where $n_{\rm D3}$ is the number of spacetime-filling D3-branes and the Euler characteristic $\chi$ is related to the Hodge numbers of $X_4$ via~\eqref{chi}.\footnote{RR tadpole cancellation conditions are usually stronger than anomaly cancellation after compactification but closely related since non-canceled RR tadpoles lead to gauge anomalies in the worldvolume of D-brane probes \cite{Aldazabal:1999nu,Aldazabal:1998mr}.} Thus, in four dimensions, the flux-less theory ($G_4\equiv 0$) requires the introduction of spacetime-filling D3-branes to cancel the tadpole for $\chi(X_4)\neq 0$. This differs from the six- or seven-dimensional case. 

Apart from anomaly/tadpole cancellation, we want to ensure that the resulting 4d theory is supersymmetric. In 6d, this requires the domain wall to be BPS as ensured by $Q^+_{G_4}=-Q^-_{G_4}$. Instead, in the four-dimensional setup, the F-term condition has to be satisfied 
\begin{equation}
    D_iW = (\partial_i + K_i) W =0\,.
\end{equation} 
Here $W$ is the superpotential, $i$ stands for all the moduli in the theory, and $K_i= \partial_i K$ is the derivative of the K\"ahler potential. Since we are mostly interested in different choices for the flux $G_4$, in the following we only consider the classical flux-induced superpotential given by~\cite{Gukov:1999ya,Haack:2001jz} 
\begin{equation}
   W= W_{\rm GVW} = \int_{X_4} \Omega_4 \wedge G_4\,,
\end{equation}
where $\Omega_4$ is the holomorphic $(4,0)$-form on the Calabi--Yau fourfold $X_4$. 

To parallel the discussion in 6d, we denote the flux-less theory with $G_4\equiv 0$ by $\mathbb{T}_0$. In this theory, the tadpole condition is satisfied by introducing $n_{\rm D3}=\frac{\chi}{24}$ spacetime-filling D3-branes. As in six dimensions, we can now ask whether, at special points in the moduli space of $\mathbb{T}_0$, another theory $\mathbb{T}_{G_4}$ can be realized that, at first sight, looks very distinct from $\mathbb{T}_0$. As a necessary condition, the on-shell value of the superpotential of the theory $\mathbb{T}_{G_4}$ has to be the same in $\mathbb{T}_0$. Since we work at the perturbative level and take $W=W_{\rm GVW}$, the superpotential of $\mathbb{T}_0$ vanishes identically. A theory $\mathbb{T}_{G_4}$ (if it exists) is thus specified by fluxes satisfying
\begin{equation}
\int_{X_4} \Omega_4 \wedge G_4 =0\,. 
\end{equation}
This condition is rather restrictive since, through $\Omega_4$, the above expression depends on the complex structure moduli $\psi^i$ of $X_4$. Since in addition we also have to preserve supersymmetry of the theory $\mathbb{T}_{G_4}$, we have to impose\footnote{Notice that for $W=0$, the supersymmetry condition is independent of the K\"ahler moduli.}
\begin{equation}\label{conditionTG4}
    W = \partial_{\psi^i} W =0 \,.
\end{equation}
 For non-zero flux, the condition~\eqref{conditionTG4} is at best satisfied along a sublocus of the complex structure moduli space of $X_4$ (if it can be satisfied at all). Let us denote this locus by $\cM_{\mathbb{T}_{G_4}}\subset \cM_{\mathbb{T}_0}$. Thus, if we tune the complex structure moduli $\psi^i$ to lie on this locus, we can nucleate the flux $G_4$ and realize the transition $\mathbb{T}_{0}\to \mathbb{T}_{G_4}$. 
 
 Within the F-/M-theory framework, such a transition is realized by nucleating an M5-brane wrapping the 4-cycle, $\Sigma_4$, dual to $G_4$. 
This, however, raises a puzzle: if the locus $\cM_{\mathbb{T}_{G_4}}$ is non-singular, the M5-brane wrapping $\Sigma_4$ is naively non-BPS.\footnote{By non-singular, we here refer to a locus at finite distance in the moduli space along which no period of $\Omega_4$ over an integral basis of calibrated four-cycles vanishes identically.} However, the tension of a BPS domain wall obtained from an M5-brane wrapping a calibrated four-cycle $\Sigma_4$ is given by the volume of this cycle, such that
\begin{equation}
    T_{M5|\Sigma_4} = \int_{\Sigma_4} \Omega_4 =  W\,,
\end{equation}
where we assumed that the cycle $\Sigma_4$ is a calibrated special Lagrangian cycle. According to the above identification, $W=0$ would imply the existence of a tensionless BPS domain wall. The locus $\cM_{\mathbb{T}_{G_4}}$ being non-singular implies that there exists an integer basis of BPS charges such that the associated periods have positive imaginary part. To achieve $W=0$ we then need to combine positive and negative charges, i.e., consider composite domain walls whose components preserve opposite supercharges. As a consequence, the composite domain wall breaks all supersymmetries and is hence not BPS. At the level of the geometry, this means that the cycle $\Sigma_4$ cannot be special Lagrangian. Thus, even though the domain wall is non-BPS, the corresponding charge $G_4\in H^4(X_4,\mathbb{Z})$ has a supersymmetric attractor locus given precisely by $\cM_{\mathbb{T}_{G_4}}$. In other words, an object that microscopically breaks all supersymmetries leads to a supersymmetric macroscopic state which is puzzling at first sight.

The puzzle can be resolved by working in analogy to the six-dimensional case discussed in Section~\ref {sec:6d}. 
To that end, we have to focus on the locus $\mathbb{T}_{G_4}$ characterized by $\partial W = W =0$ (assuming it exists for a given flux). For M-theory on Calabi--Yau fourfolds, the quantization condition reads \cite{Witten:1996md}
\begin{equation}
    G_4 + \frac{c_2(X_4)}{2} \in H^4(X_4,\mathbb{Z})\,. 
\end{equation}
For simplicity, we focus on Calabi-Yau fourfolds with an even second Chern class, for which we can choose an integer-quantized $G_4$-flux to satisfy the quantization condition. This restriction allows us to focus on primitive fluxes satisfying $J_{X_4}\wedge G_4=0$, for $J_{X_4}$ the K\"ahler form on $X_4$, and hence not break Poincar\'e invariance in the 4d F-theory lift of the M-theory setup. The four-form flux $G_4$ is then an element of the horizontal cohomology
\begin{equation}
    H^4(X_4,\mathbb{C})_{\rm hor.} = H^{4,0}(X_4) \oplus H^{3,1}(X_4) \oplus H^{2,2}(X_4)_{\rm hor.} \oplus H^{1,3}(X_4) \oplus H^{0,4}(X_4)\,. 
\end{equation}
At a generic point in the complex structure moduli space, an integer-quantized, primitive flux $G_4$ has components in each of the above spaces that make up $H^4(X_4,\mathbb{C})_{\rm hor.}$. At the special locus along which \eqref{conditionTG4} is satisfied, the flux instead only has components along $H^{2,2}(X_4)$~\cite{Becker:1996gj,Gukov:1999ya,Haack:2001jz}. In other words, there is an integer, primitive four-form that is of pure $(2,2)$-type. Such a $(2,2)$-form defines a Hodge class, and the locus in the moduli space along which the form is of Hodge-type $(2,2)$ is referred to as its Hodge locus. Such Hodge loci are very special and require the periods of the fourfold to have a particularly simple form such that the condition~\eqref{conditionTG4} admits a solution. 

The problem of finding solutions to~\eqref{conditionTG4} simplifies if, locally, the periods can be expressed as polynomials without any exponential correction terms. 
This happens, for example, along loci in the complex structure moduli space of $X_4$ at which the full Picard--Fuchs system of $X_4$ reduces to the Picard--Fuchs system of K3. Such loci were investigated in~\cite{Grimm:2024fip} for the fourfold analogue of the Hulek--Verrill threefold~\cite{hulek2005modularityrigidnonrigidcalabiyau} first considered in~\cite{Jockers:2023zzi} for which exact solutions to~\eqref{conditionTG4} were found.\footnote{In particular, \cite{Grimm:2024fip} identify an exact supersymmetric $W=0$ vacuum for a $G_4$-flux that cancels the tadpole exactly without introducing additional spacetime-filling D3-branes.} Moreover, \cite{Grimm:2024fip} argues in general, the condition $W=\partial W=0$ can only be solved over the integers at symmetric points in the moduli space where the relevant functions are algebraic as opposed to transcendental. This happens, for example, if the relevant periods are associated to a K3 instead of a genuine Calabi--Yau fourfold.\footnote{Enhanced supersymmetry is not the only way to realize $W=0$ vacua. Another possibility is to realize such supersymmetric flux vacua at special symmetric loci in the moduli space, such as Fermat points~\cite{Braun:2020jrx}, along which, however, certain integer periods vanish, thus automatically leading to BPS M5-branes with vanishing tension in string units.}

Such exact solutions to the condition~\eqref{conditionTG4} are hence examples for the theories $\mathbb{T}_{G_4}$ that we consider in this work. The space $\cM_{\mathbb{T}_{G_4}}$ in this case corresponds to the Hodge locus for $G_4$. To summarize, we can consider the locus $\mathbb{T}_{G_4}$ with the following properties:
\begin{enumerate}
    \item Along $\cM_{\mathbb{T}_{G_4}}$ there is an additional integer 4-form of type $(2,2)$. 
    \item The local periods along $\cM_{\mathbb{T}_{G_4}}$ can be expressed as the periods of K3. 
    \item The transition $\mathbb{T}_{G_4}\rightarrow \mathbb{T}_{0}$ corresponds to removing $G_4$-flux.
\end{enumerate}
These three properties can be interpreted as the analogue of properties 2., 3. and 5. of the locus $\mathbb{T}_2$ of the six-dimensional $\cN=(1,0)$ setup summarized at the end of Section~\ref{sec:6d}: the appearance of an additional holomorphic curve (i.e., a two form of type (1,1)), the local hyperK\"ahler geometry along $\cM_{\mathbb{T}_2}$ translating into an enhancement of supersymmetry and the removal of $G_4$-flux on one of the 9-branes at the end of the M-theory interval.

We now give a physical argument supporting the claim in~\cite{Grimm:2024fip}, i.e., argue why supersymmetric flux configurations with $W=0$ can, in particular, be realized along loci in the complex structure moduli space with enhanced supersymmetry. We therefore recall that the M5-brane on the four-cycle Poincar\'e dual to $G_4$ has a supersymmetric attractor point at $\cM_{G_4} \subset \cM_{\rm c.s.}(X_4)$ even though the system of M5-branes does not preserve any of the four supercharges of the underlying 4d $\cN=1$ supersymmetric theory. However, if supersymmetry is enhanced, the domain wall can be protected by a BPS condition in the algebra of the enhanced supersymmetry. Let us denote the four supercharges of the original 4d $\cN=1$ theory by 
\begin{equation}
   \text{universal 4d}\;\cN=1\text{ supercharges}:\qquad  Q_\alpha, Q_{\dot{\alpha}}\,, \qquad \alpha, \dot{\alpha}=1,2\,.
\end{equation}
We now assume that there exists a locus in $\cM_{\rm c.s.}(X_4)$ along which we can define four additional supercharges $\tilde{Q}_\alpha, \tilde{Q}_{\dot \alpha}$ that are unbroken in the vicinity of the Poincar\'e dual of $G_4$ such that the local $\cN=2$ supercharges 
\begin{equation}
    \text{local 4d}\;\text{ supercharges}:\qquad (\tilde{Q}_\alpha, \tilde{Q}_{\dot \alpha})\,,\qquad \alpha,\dot \alpha=1,2\,,
\end{equation} 
together with the universal supercharges form an $\cN=2$ algebra in four dimensions. For such a local supersymmetry enhancement to occur, the geometry in the vicinity of ${\rm P.D.}(G_4)$ has to have reduced holonomy, i.e., involve, for example, a K3 surface. This is consistent with the results of \cite{Grimm:2024fip}. The domain wall corresponding to the M5-brane on ${\rm P.D.}(G_4)$ breaks the supercharges $Q_{\alpha,\dot \alpha}$ while preserving $\tilde{Q}_{\alpha, \dot\alpha}$ and is hence a $\frac12$-BPS state of the enhanced $\cN=2$ superalgebra. For this reason, its tension is a BPS-protected quantity. If the locus $\cM_{\mathbb{T}_{G_4}}$ is contained in the locus with enhanced supersymmetry, then the tension of the domain wall is indeed vanishing since it is a BPS state of the enhanced supersymmetry algebra. 

Thus, the domain wall interpolating between the theories $\mathbb{T}_{G_4}\to \mathbb{T}_0$ has zero tension such that we can move between the two theories at no energy cost. As a consequence, the theories $\mathbb{T}_{G_4}$ and $\mathbb{T}_0$ are realized in the same moduli space. Notice that, depending on the codimension of $\mathcal{M}_{\mathbb{T}_{G_4}}\subset \cM_{\rm c.s.}(X_4)$, the naive dimension of the moduli space of $\mathbb{T}_{G_4}$ can be significantly smaller than the dimension of the moduli space of $\mathbb{T}_0$. Still, since we can move from $\mathbb{T}_{G_4}\to \mathbb{T}_0$ at zero energy cost, the actual moduli space of $\mathbb{T}_{G_4}$ is enhanced. The process bringing us from $\mathbb{T}_{0}\to \mathbb{T}_{G_4}$ is non-perturbative in nature as it requires the nucleation of an M5-brane domain wall. Therefore, this transition is not visible from the perturbative string theory perspective. This highlights that in theories with minimal supersymmetry, the perturbative perspective might not capture all the physical processes. 

It is worth emphasizing again the similarities with the setup in six dimensions, which were discussed in section~\ref{sec:6d}. In the 6d case, we saw that at generic points in $\cM_{\rm c.s.}(X_3)$ the curve $h=f-g$ is not holomorphic, such that the combination of strings $\mathtt{S}_f$ and $-\mathtt{S}_g$ is not BPS and breaks all eight supercharges. Still, along the special locus $\cM_{\mathbb{T}_2}$, the supersymmetry locally gets enhanced due to the local hyperK\"ahler geometry of the base $B_2$. Along $\cM_{\mathbb{T}_2}$, the curve $h$ is holomorphic, and thus yields a BPS string when wrapped by a D3-brane.\footnote{Moreover, the 4d situation discussed in \ref{sec:mojodojo} is not different. From the perspective of F-theory on $\widehat{X}_4$, the curve $C'-C''$ is not holomorphic. However, in the phases~\hyperref[phaseII,III]{II and III} wrapping a D3-brane on this curve yields the 4d string with enhanced supersymmetry.}
The situation in the 4d setup discussed here is perfectly analogous, just that the extended objects arise from M5-branes wrapping four-cycles leading to domain walls in 4d or strings in 3d M-theory compactifications on $X_4$.

The analogy of the 4d flux case with the setups discussed in sections~\ref{sec:6d} and \ref{sec:mojodojo} is summarized in the following table
\begin{center}
    \begin{tabular}{|c||c|c|c|}\hline 
         & 6d: $\mathbb{T}_2\to\mathbb{T}_0$ &4d flop: $X_4\rightarrow\widehat{X}_4$ &  4d flux: $\mathbb{T}_{G_4}\to \mathbb{T}_0$ \\\hline \hline 
        SUSY enh. locus & $\cM_{\mathbb{T}_2}\subset \cM_{\rm c.s.}(X_3)$ & $\mathcal{M}_{X_4}\subset\mathcal{M}_{\widehat{X}_4}$ & $\cM_{\mathbb{T}_{G_4}}\subset \cM_{\rm c.s.}(X_4)$    \\ \hline
        Extra BPS state & D3 on $h=f-g$& D3 on $C'-C''$ & M5 on $\Sigma_4$ \\ \hline 
        Local Holonomy & $SU(2)$ & $SU(3)$ & $SU(2)$ \\\hline
        \end{tabular}
\end{center}
As mentioned above, there is one difference to the 6d case, namely that the change in the flux has to be compensated by the presence of spacetime-filling D3/M2-branes to cancel the tadpole induced by the curvature of $X_4$. Thus, the theory $\mathbb{T}_{0}$ has 
\begin{equation} 
\delta n_{\rm D3/M2}=\frac12 \int G_4\wedge G_4
\end{equation} 
additional D3/M2-branes compared to $\mathbb{T}_{G_4}$. The position of each spacetime-filling D3/M2-brane along  $X_4$ contributes real scalar degrees of freedom that we denote by $\mathbf{x}^{(\alpha)}$, $\alpha=1,\dots, n_{D3/M2}$. Thus, these D3/M2-branes contribute an additional sector to the moduli space corresponding to their position moduli. We discuss local transitions involving D3-brane position moduli in section~\ref{ssec:D3moduli}.

\paragraph{Supersymmetry Breaking Effects. }
In the scenario discussed above, we encountered enhanced supersymmetry along a special locus in the complex structure moduli space of $X_4$. Still, even though supersymmetry is enhanced in specific subsectors of the theory, globally we still have only 4d $\cN=1$ supersymmetry. A crucial question is then how the effects that globally break $\cN=2\to \cN=1$ supersymmetry enter. This is analogous to our discussion in Section~\ref {ssec:global} of global supersymmetry breaking affecting the subsector with enhanced supersymmetry.

The key insight is that so far, we have treated the different sectors of the moduli space as completely decoupled. However, from the perspective of the 4d $\cN=1$ theory, this decoupling is unexpected since all scalar fields reside in chiral multiplets and hence form part of a single scalar field space. In particular, in the discussion of the $W=0$ flux vacua above, we considered the complex structure sector as completely decoupled from the K\"ahler sector. This decoupling stems from the classical form of the 4d $\cN=1$ K\"ahler potential 
\begin{equation}
    K= -2\log \left(\int_{B_3} J_{B_3}^3\right) - \log\left(\int_{X_4}\Omega\wedge \bar\Omega\right)\,. 
\end{equation}
While this K\"ahler potential is a valid approximation in the limit $V_{B_3} =\frac16 \int_{B_3}J_{B_3}^3 \to \infty$ limit, quantum corrections spoil this factorization at finite $V_{ B_3}$. Corrections to the K\"ahler potential that manifestly break the $\cN=2$ structure are of the form 
\begin{equation}\label{deltaK}
    \delta K = -\log \left[V_{B_3}^{-1} \left(\mathcal{A}_0(\psi_i,\bar\psi_i) + \sum_{\mathbf{k}=0} \mathcal{A}_\mathbf{k}(\psi_i,\bar{\psi_i}) e^{-2\pi T_{\mathbf{k}}} + \text{c.c.}\right) \right]\,,
\end{equation}
where the $\psi_i$ are complex structure moduli and the $T_i$ are K\"ahler moduli. Here $\mathbf{k}$ runs over suitable elements in $H^2(X_4)$ but, since the K\"ahler potential in $\cN=1$ theories is not BPS, we do not have to restrict to BPS instantons in the above expression. As is evident from the above expression, these corrections mix the complex structure and K\"ahler sector and are hence an indication that away from certain subsectors the theory only preserves $\cN=1$ supersymmetry.


In addition to non-perturbative corrections to the K\"ahler potential, there could also be contributions to the superpotential at finite $T_k$ coming from BPS D3-brane instantons wrapping holomorphic divisors inside $B_3$. These could induce a runaway potential towards the large volume regime $V_{B3}\to \infty$. As discussed above, in this regime, the K\"ahler and complex structure sector effectively decouples such that our analysis of the supersymmetry enhancement in the complex structure sector is unaffected even in the presence of such a non-perturbative superpotential. In other words, non-perturbative corrections to the superpotential may dynamically force us into the regime of moduli space in which our analysis of the loci featuring enhanced supersymmetry has been carried out. This is similar to the discussion of the non-perturbative superpotential generated in the bulk at the end of section~\ref{sec:mojodojo}.\newline

To summarize, in this section, we considered subsectors with enhanced supersymmetry by considering the supersymmetric vacua with $W_{\rm GVW}=0$. Since the scalar potential vanishes identically in these subsectors, the theories resemble $\cN=2$ theories, as they feature an exact moduli space. Therefore, the dynamics of the theory indeed behave as expected for enhanced supersymmetry. However, the global supersymmetry breaking becomes visible at the kinematic level since the K\"ahler potential breaks the factorization between the different sectors that would be decoupled in an actual 4d $\cN=2$ theory.

\section{Heterotic Dual Perspective}\label{sec:heterotic}

In sections~\ref{sec:mojodojo} and \ref{sec:flux}, we encountered two cases of transitions in 4d $\cN=1$ theories based on F-theory compactifications that effectively augment the size of the naive moduli space of a given theory. Still, the two examples, i.e., the flux and the flop transitions, appear to be quite different in their microscopic descriptions. In this section, we employ a dual perspective that provides a unified description of these two transitions in terms of a single type of transition within a dual formulation of the theory. As a disclaimer, we note that in this section we do not work out the dual formulation in full detail. Instead, this section is supposed to provide a qualitative picture of how the transitions discussed in the previous sections can be unified in a dual formulation of the theory.

To that end, we restrict to a certain class of compactification manifolds for the original F-theory formulation of the theory. Concretely, we assume that the elliptically-fibered Calabi--Yau fourfold $X_4$ has a \emph{compatible} K3-fibration. Importantly, this implies that the base $B_3$ is itself rationally fibred 
\begin{equation}\label{fibrationB3}
    \mathbb{P}^1\stackrel{\rho}{\hookrightarrow} B_3 \to B_2 \,,
\end{equation}
where $B_2$ is a K\"ahler surface. This is a necessary condition for the theory to have a heterotic dual~\cite{Morrison:1996na,Morrison:1996pp,Lee:2019jan}. Let us assume that such a heterotic dual exists. In this case, the heterotic string is compactified on an elliptically fibered Calabi--Yau threefold
\begin{equation}
    \cE \stackrel{p}{\hookrightarrow}X_H\to B_2 \,,
\end{equation}
over the same K\"ahler surface $B_2$ appearing in \eqref{fibrationB3}. Additionally, a gauge bundle on $X_3$ is required to satisfy the heterotic Bianchi identity. In the following, we restrict to the heterotic $E_8\times E_8$ string. On the F-theory side, the details of the gauge bundle are encoded in the twist of the $\mathbb{P}^1$ over $B_2$ given by the projection $\rho$ as well as in the $G_4$-flux~\cite{Friedman:1997yq,Nazi:1998bva}. 

In addition to a non-trivial gauge bundle, the heterotic Bianchi identity can also be canceled by spacetime-filling NS5-branes wrapping curves inside $X_H$ as 
\begin{equation}
    [{\rm NS5}] = c_2\left(\Omega_{X_H}\right) - c_2\left(V\right)\,. \label{bianchi}
\end{equation}
Here $[{\rm NS5}]\in H_2(X_H)$ is the class of the curve wrapped by the NS5-branes inside $X_3$.  
Spacetime-filling NS5-branes in the heterotic compactification on $X_H$ map to different objects in F-theory depending on the curve class $[{\rm NS5}]\in H_2(X_H)$ is wrapped by the heterotic NS5-brane. We can distinguish two cases: $i)$ the NS5-branes wraps the elliptic fiber of $X_H$, i.e. $[{\rm NS5}]=n_{5}\cE$ or $ii)$ the NS5-branes wraps a curve in the base $B_2$ of $X_H$, i.e., $[{\rm NS5}]\in H_2(B_2)$: 
\begin{itemize}
    \item[$i)$] The F-theory dual of a heterotic NS5-brane wrapping $n_5$-times the fiber $\cE$ has been identified in \cite{Nazi:1998bva} as spacetime-filling D3-branes. Recall that $G_4$-flux on the F-theory side maps to the heterotic gauge bundle data. Therefore, a transition in F-theory replacing non-zero $G_4$-flux by spacetime-filling D3-branes maps on the heterotic dual side to a transition replacing non-trivial gauge bundle data with NS5-branes wrapping the elliptic fiber of $X_H$. 
    \item [$ii)$] Heterotic NS5-branes wrapping curves in the base, $B_2$ of $X_H$, are completely geometrized in F-theory~\cite{Haghighat:2014pva}: Let $C_b=[{\rm NS5}]\in H_2(B_2)$ be the base curve wrapped once by the NS5-branes in the heterotic duality frame. Concretely, from the perspective of the base $B_3$ of the F-theory compactification, the NS-brane wrapped on $C_b\in H_2(B_2)$ manifests in a degeneration of the fiber $\mathbb{P}^1$ of $B_3$ over $C_b$. In other words, over $C_b$ the fibral $\mathbb{P}^1$ degenerates into the union of two (or more) rational curves intersecting over points.
\end{itemize}

Let us focus on the second case and consider a curve $C_b\in H_2(B_2)$ and investigate the transition between the heterotic bundle data and NS5-branes wrapping $C_b$. From~\cite{Friedman:1997yq} (see \cite{Weigand:2010wm} for a review), we recall that in case the bundle is given as a spectral line bundle, the projection of the heterotic line bundle to the base $B_2$ can be expressed in terms of the geometry of the F-theory compactification manifold. Concretely, the class of the heterotic $E_8\times E_8$ gauge bundles projected to the base $B_2$ is determined by the geometry of the dual F-theory compactification via 
\begin{equation}
    \eta_1 = p_*(\lambda(V_1)) = 6c_1(B_2) - c_1(\cT)\,,\qquad \eta_2=p_*(\lambda(V_2)) = 6 c_1(B_2) + c_1(\cT) \,. 
\end{equation}
Here $\lambda(V_i)$ denotes the class of the vector bundle in the $i$-th gauge factor and $\cT$ is the twist bundle of the fibration $\rho$ on the F-theory side. 

We now analyze the transition discussed in section~\ref{sec:mojodojo} from this heterotic dual perspective. To that end, we consider a curve $C_b\in H_2(B_2)$ for which the embedding into $B_3$ yields a curve $C_0\in H_2(B_3)$ with normal bundle
\begin{equation}\label{normalbundlehet}
    \cN_{C_0/B_3} = \cO(-1)\oplus \cO(-1)\,. 
\end{equation} 
Let us denote by $S_\pm$ the plus/minus section of $\rho$ satisfying 
\begin{equation}
    S_+= S_- + \rho^*(c_1(\cT)) \,,\qquad S_-\cdot_{B_3} S_+ =0\,. 
\end{equation}
Furthermore, let $\rho^*(C_b)$ be the vertical divisor over $C_b$ w.r.t. the fibration $\rho$. Then we can write 
\begin{equation}
    C_0 = S_- \cdot_{B_3} \rho^* (C_b)\,.
\end{equation}
Since $C_0$ is effective, the normal bundle \eqref{normalbundlehet} of $C_0$ requires  
\begin{align}\label{c1TCb}
    C_0\cdot_{B_3} S_-&= S_-\cdot_{B_3} S_-\cdot_{B_3} \rho^*(C_b)= -c_1(\cT)\cdot_{B_2} C_b = -1 \,,\\
    C_0 \cdot_{B_3} \rho^*(C_b) &= S_-\cdot_{B_3} \rho^*(C_b)\cdot_{B_3} \rho^*(C_b)=C_b\cdot_{B_2} C_b=-1 \,.
\end{align}
In other words, the curve $C_b$ has to have self-intersection $-1$ inside $B_2$, and it has to intersect the first Chern class of the twist bundle $\cT$ once. Using that for flat, rational fibrations with twist bundle $\cT$ over a base $B_2$, the Chern class of the total space $B_3$ is given by 
\begin{equation}
    c_1(B_3) = 2S_-+\rho^*(c_1(B_2)) + \rho^*(c_1(\cT)) \,,
\end{equation}
we find 
\begin{equation}
c_1(B_3)\cdot_{B_3} C_0 =0 \quad \Leftrightarrow \quad c_1(B_2)\cdot_{B_2} C_b = c_1(\cT)\cdot_{B_2} C_b\,.     
\end{equation}
This identity can be used to determine the component of the class of the respective gauge bundles along the curve $C_0$ as 
\begin{equation}
    \eta_1\cdot_{B_2} C_b = 5c_1(B_2)\cdot_{B_2}C_b = 5\,,\qquad \eta_2\cdot_{B_2} C_b=7c_1(B_2)\cdot_{B_2} C_b=7\,.
\end{equation}
In the heterotic/M-theory picture, this means that there is effectively $\pm 1$ unit of flux along $C_b$ in the two 9-branes at the end of the interval.\footnote{We notice that this is the analogue of the heterotic setup discussed in section~\ref{sec:6d}. To obtain the topological data of the gauge bundle in this six-dimensional case, we have to replace $c_1(B_2)\cdot_{B_2}C_b \to c_1(B_1=\mathbb{P}^1)=2$ because in 6d the base $B_1$ of the elliptic K3 on the heterotic side is simply a $\mathbb{P}^1$. With these replacements, the above expression leads to the instanton embedding $(10,14)$ in 6d.} 

\begin{figure}[!t]
    \centering
\includegraphics[width=0.5\linewidth]{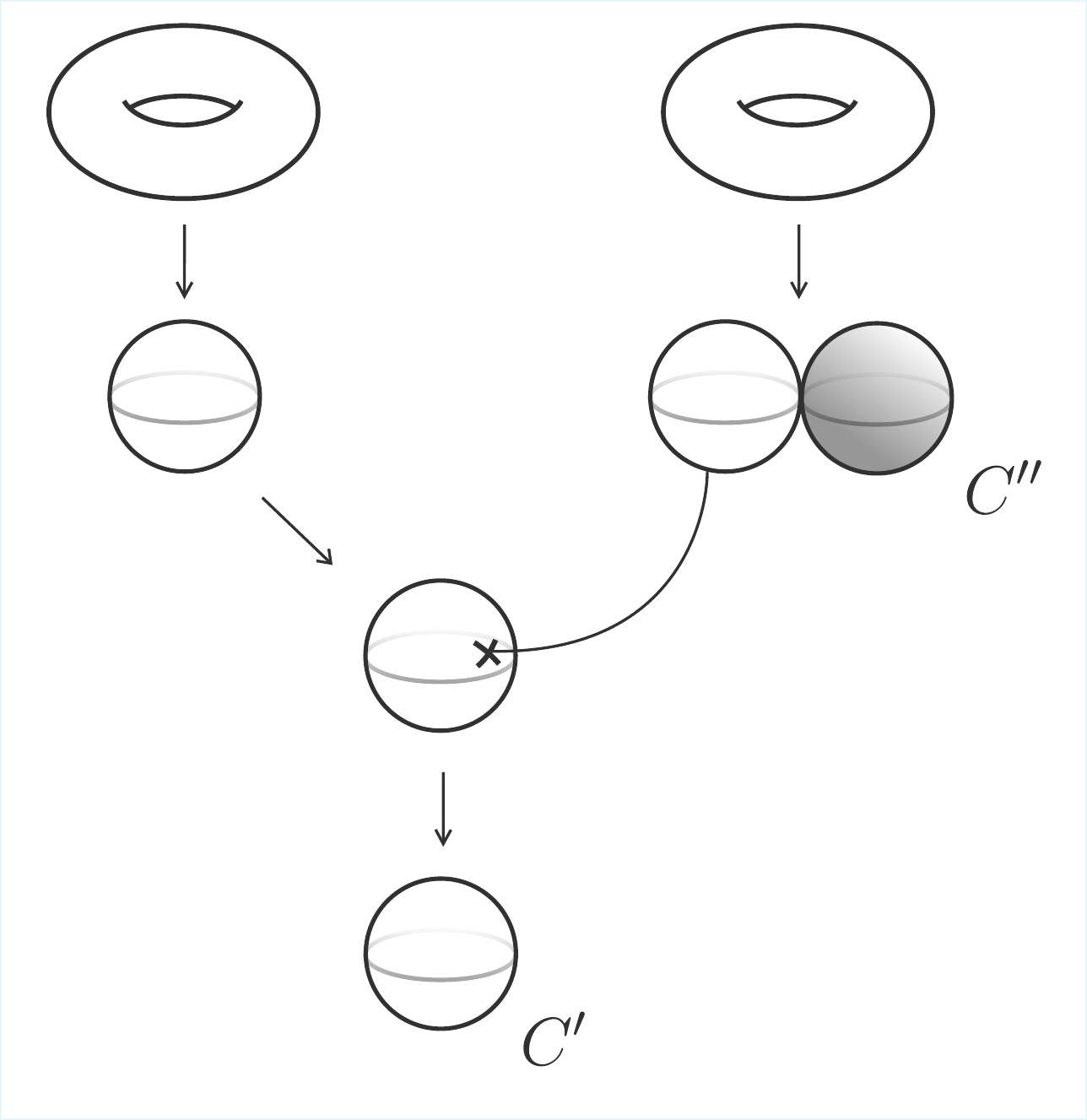}
    \caption{A sketch of the fibration structure of the Calabi Yau 4-fold $\widehat{X}_4$ considered in this chapter $T^2\to \mathbb{P}^1_F \to B_2 =\mathbb{F}_1$. The fiber $\mathbb{P}^1_f$ degenerates over $C'$ in the union of two rational curves intersecting over a point corresponding to a blow-up of the base curve of $\mathbb{F}_1$.}
    \label{fig: fibration}
\end{figure}

In the F-theory duality frame, the transition described in section~\ref{sec:mojodojo} corresponds to blowing up the curve $C_0\subset B_3$. This blow-up introduces the exceptional divisor $E\simeq \mathbb{P}^1\times \mathbb{P}^1$. Whereas one of the $\mathbb{P}^1$s is the proper transform of $C_0$, the other one is an additional fibral $\mathbb{P}^1_f$ in the split of the $\mathbb{P}^1$ fiber over $C_0$ into the union of two $\mathbb{P}^1$s. We identify the proper transform of $C_0$ and the additional fibral curve as the respective curves $C'$ and $C''$ introduced in section~\ref{sec:mojodojo}. This setup is represented in Fig.~\ref{fig: fibration}. As reviewed above, the presence of one additional fibral curve over $C'$ translates into a heterotic NS5-brane wrapping $C_b$ on the heterotic dual side. 

The transition on the F-theory side thus corresponds to taking out one unit of NS5-brane charge along $C_b$ from the gauge bundle $V_2$. Thus, its class along the base transforms as 
\begin{equation}
    \eta_2 \to \tilde\eta_2 \quad \text{with} \quad \tilde{\eta}_2\cdot_{B_2} C_b = \eta_2\cdot_{B_2} C_b -1 =6\,. 
\end{equation}
Thus, we completely remove the 5-brane charge from the visible 9-brane in the Horava--Witten picture and instead nucleate a 5-brane on the curve $C_b$. Even though the nucleation of a spacetime-filling 5-brane might seem to be a rather violent process, this is simply the 4d analogue of small instanton transitions in heterotic compactifications on K3~\cite{Witten:1995gx}. Since these transitions correspond to extremal transitions in the moduli space of the 6d $\cN=(1,0)$ theory, the nucleation of the spacetime-filling NS5-brane corresponds to a finite (and in fact zero) energy process. From this point of view our proposal is hence simply a variant of a well-established process in string theory. For the transition discussed in section~\ref{sec:6d}, the relevant strings $\mathtt{S}_f$ and $\mathtt{S}_g$ are critical heterotic strings that, at the special locus, $\cM_{\mathbb{T}_2}$, in the moduli space form a bound state. This bound state, the M-string obtained by wrapping a D3-brane on a (-2)-curve in the base of the F-theory threefold, does not have an analogue in $d>6$ dimensions. This is closely related to the fact that the heterotic-heterotic duality exchanging the two critical heterotic strings in the $\mathbb{T}_0$-theory only occurs $d\leq 6$ dimensions. 

In our four-dimensional case, we encounter yet another string $\mathtt{S}_0$, namely the D3-brane wrapping the curve $C_0$ with normal bundle \eqref{normalbundlehet}. This kind of string has no analogue in six dimensions or higher.  In contrast, the strings $\mathtt{S}'$ and $\mathtt{S}''$ involved in the transition are four-dimensional cousins~\cite{Lee:2020gvu} of 6d E-strings \cite{Klemm:1996hh,Witten:1996qb,Lerche:1996ni} since 
\begin{equation}
    \mathcal{N}_{C'/\hat B_3} = \mathcal{O}(-1)\oplus \mathcal{O}(0) = \cN_{C''/\hat B_3}\,.
\end{equation}
These strings are closely related to NS5-branes in the dual heterotic compactification~\cite{Ganor:1996mu,Haghighat:2014pva}. The worldsheet theory of the strings $\mathtt{S}'$ and $\mathtt{S}''$ thus resembles the worldsheet theory of a D3-brane on a (-1)-curve in 6d, i.e., an E-string. At generic points in the K\"ahler moduli space of $\widehat{B}_3$, the combination of curves $C'-C''$ does not yield a BPS string. In other words, the two E-strings do not yield a bound state if wrapped with opposite orientation. Instead, along $\cM_{B_3}\subset \cM_{\widehat{B}_3}$, the string $\mathtt{S}_0$ exists and is BPS. This string can be viewed as a bound state of the string $\mathtt{S}'$ and $-\mathtt{S}''$. This is the 4d analogue of the string $\mathtt{S}_h$ being a bound state of $\mathtt{S}_f$ and $\mathtt{S}_g$ that only exists along the special locus $\cM_{\mathbb{T}_2}$ in the 6d moduli space. The difference between the 4d and 6d cases is that in 6d, the strings $\mathtt{S}_f$ and $-\mathtt{S}_g$ are (mutually non-local) heterotic strings, whereas in 4d the strings $\mathtt{S}'$ and $\mathtt{S}''$ are E-like strings. For this reason, it is not surprising that in 4d the transition involves spacetime-filling heterotic NS5-branes wrapping curves in the heterotic $X_H$, which are indeed associated with E-strings in F-theory. The key properties of the 4d transition and their 6d analogues are summarized in Table~\ref{tab:6dvs4d}. 
\begin{table}[!t]
\begin{center}
    \begin{tabular}{|c|c|c|}\hline 
         & 6d local SUSY enhancement &  4d local SUSY enhancement\\\hline \hline
       Curve $C_0$ & $\cN_{C_0/B_2}=\cO(-2)$ & $\cN_{C_0/B_3}=\cO(-1)\oplus \cO(-1)$    \\ \hline
        Massless d.o.f. & $\cN=(1,0)$ tensor mult. & $\cN=1$ chiral mult. \\ \hline 
       Massive FT sector & 6d $\cN=(2,0)$ SQFT  & massive 4d $\cN=2$ vector mult. \\\hline
       Additional modulus & complex structure + axions & blow-up of $C_0$ + axions \\\hline 
       New curves& $\cN_{f,g/B_2}=\cO$ & $\cN_{C',C''/B_3}=\cO(-1) \oplus \cO$ 
       \\\hline 
       D3-brane strings  & Heterotic Strings $(\mathtt{S}_f, \mathtt{S}_g)$&  E-Strings ($\mathtt{S}',\mathtt{S}''$)\\ \hline  
       \end{tabular}
\end{center}
\caption{Summary of the key players in the transition associated with local supersymmetry enhancement in 4d and their counterparts in 6d.\label{tab:6dvs4d}}
\end{table}

From the above discussion, we see that the condition $C_0\cdot_{B_3}\bar{K}_{B_3}=0$ is central for the transition to work on the heterotic side: only in this case, a single NS5-brane has to be nucleated to remove the flux from the visible 9-brane in the heterotic dual theory. If instead we considered a curve with $C_0\cdot_{B_3}\bar{K}_{B_3}=-1$, we would need two NS5-branes, which is not necessarily possible in a supersymmetric way. On the other hand, 
\begin{equation}
    C_0\cdot_{B_3}\bar{K}_{B_3}>0 \quad \Rightarrow \quad \eta_2\cdot_{B_2}C_0 <7 \,,
\end{equation} 
such that in this case there is not enough flux in the visible 9-brane to be removed by an integer amount of NS5-branes. 

So far, we have not addressed the global structure of the heterotic compactification. In section~\ref{ssec:global}, we saw that in the dual F-theory picture, there is a striking difference between the blow-up in a non-compact geometry and its embedding in a compact geometry. In particular, in the global embedding, the transition $X_4\to \widehat{X}_4$ is associated with a reduction of the number of complex structure moduli and the number of spacetime-filling D3-branes, see~\eqref{eq:deltand3} and~\eqref{eq:deltah31}. The reduction in the number of D3-branes maps in the heterotic theory to a reduction of NS5-branes wrapping the fiber of $X_H$. Moreover, the reduction in the number of complex structure moduli can also be associated with a change in the number of 5-branes in the heterotic theory. To illustrate this, let us reconsider the blow-up in the geometry described above. As in section~\ref{sec:mojodojo}, we denote the proper transformation of $C_0\subset B_3$ by $C'\subset \widehat{B}_3$, i.e.,
\begin{equation}
    C_0 = \phi_*(C')\,,
\end{equation}
where, as in~\eqref{C0Cprime}, we introduced the birational morphism $\phi:\widehat{B}_3\to B_3$. After the blow-up, the base $\widehat{B}_3$ remains rationally fibred over the same base as before 
\begin{equation}
    \mathbb{P}^1_f \stackrel{\widehat{\rho}}{\hookrightarrow} \widehat{B}_3 \to B_2\,,
\end{equation}
with the only difference that over the curve $C_b\in H_2(B_2)$ the rational fiber splits into a union of two $\mathbb{P}^1$s. We can thus write
\begin{equation}
    C'= \widehat{S}_- \cdot_{B_3} \widehat{\rho}^*(C_b)\,. 
\end{equation}
However, compared to the original curve $C_0\subset B_3$, the normal bundle of $C'$ is 
\begin{equation}
    \cN_{C'/B_3} = \cO(-1)\oplus \cO(0)\,. 
\end{equation}
By \eqref{c1TCb}, after the blow-up this requires
\begin{equation}
    c_1(\widehat{\cT})\cdot_{B_2} C_b = 0\,,
\end{equation}
where $\widehat{\cT}$ is the twist bundle of the rational fiber in $\widehat{B}_3$. Thus, through the blow-up, the twist of the rational fibration changes as 
\begin{equation}
    c_1(\cT)\cdot_{B_2} C_b = 1 \quad \longrightarrow \quad c_1(\widehat{\cT}) \cdot_{B_2} C_b=0\,. 
\end{equation}
In other words, the rational fibration becomes trivial in a vicinity of $C_b$ after the blow-up. This is consistent with the rational fiber splitting into two components 
\begin{equation}
    \mathbb{P}^1_f \to \mathbb{P}^1_1 +\mathbb{P}^1_2\,,
\end{equation}
that remain constant over the curve $C_b$. Still, the information contained in the non-trivial fibration over $C_b$ before the blow-up cannot be lost. The parameters associated with the non-trivial fibration can be identified with complex structure moduli of $X_4$. To understand how these have to be tuned after the blow-up, let us consider the simple case with $B_2=\mathbb{F}_1$. In this case, $B_2$ is itself rationally fibred over $C_b\simeq \mathbb{P}^1$. Indeed, for $B_2 = \mathbb{F}_1$, the self-intersection of the base is 
\begin{equation}
    C_b\cdot_{\mathbb{F}_1} C_b =-1\,,
\end{equation}
as required for the normal bundle of $C_0$ before the blow-up to be given by \eqref{normalbundlehet}. From the perspective of $\mathbb{F}_1$, the curve $C_b$ can be identified as the zero-section $\CZ_-:C_b \to \mathbb{P}^1_B$ where 
\begin{equation}
    \mathbb{P}^1_B \hookrightarrow \mathbb{F}_1 \to C_b\,,
\end{equation}
is the fibration structure of $B_2=\mathbb{F}_1$. The non-compact model can now be obtained by replacing the fiber with a complex plane 
\begin{equation}
    \mathbb{P}^1_B \to \mathbb{C}\,. 
\end{equation}
Prior to the blow-up, the fibration of $\mathbb{P}^1_f$ is non-trivial in the vicinity of $\CZ_-$. Instead, after the blow-up, this fibration becomes trivial in the vicinity of $\CZ_-$. In the non-compact case, all the information of the initially non-trivial fibration can be sent to infinity and thus be completely decoupled from the local physics around $\CZ_-$. 

In the compact case, one cannot send all information of the non-triviality of the rational fibration of $B_3$ over $\mathbb{F}_1$ to infinity, but only to the ``section at infinity'' $\CZ_+$, i.e., the point added in the one-point compactification that underlies $\mathbb{P}^1$. Within $\mathbb{F}_1$, the zero section and the section at infinity satisfy the important property
\begin{equation}
    \CZ_-\cdot_{\mathbb{F}_1 }\CZ_+ =0\,. 
\end{equation}
Thus, classically, the local physics associated with the section $\CZ_-$ is independent of the physics at $\CZ_+$. In the compact model, the information about the $\mathbb{P}^1_f$ fibration is localized at $\CZ_+$ since it is trivial over $\CZ_-$. In other words, the complex structure deformations of the geometry before the blow-up are tuned in such a way that the non-trivial twist of $\mathbb{P}^1_f$ localizes to $\CZ_+$. Thus, the complex structure moduli are tuned to engineer a new interacting field theory subsector at $\CZ_+$ containing all the information previously held in the non-trivial twist of $\mathbb{P}^1_f$ over $\mathbb{F}_1$. 

From the heterotic perspective, such a field theory subsector is typically associated with NS5-branes, which have to cluster around $\CZ_+$. This can thus be interpreted as NS5-branes wrapping the elliptic fiber of $\CE$ of $X_H$ localized at $\CZ_+$. The non-compact limit for $\mathbb{F}_1$ maps on the heterotic side to a limit where, again, the fiber of $\mathbb{F}_1$ is replaced by a complex plane. However, on the heterotic side, we furthermore take the 4d string coupling $g_4\to \infty$. It is defined as 
\begin{equation}
    \frac{1}{g_4^2 }= \frac{1}{g_{10}^2} \text{vol}(X_H)\, M_{\rm H}^6\,,
\end{equation}
where $g_{10}$ is the 10d heterotic string coupling and $M_{\rm H}$ the heterotic string scale. The local F-theory model maps to a weakly-coupled heterotic string on a Calabi--Yau threefold with a non-compact base, i.e., 
\begin{equation}
    (\mathbb{P}^1_B \to \mathbb{C})_F \quad \longleftrightarrow \quad (\mathbb{P}^1_B\to \mathbb{C}, g_4 \to \infty)_H\,. 
\end{equation}
In this limit, the physics of NS5-branes wrapping $\cE$ decouples $i)$ because of the weak-coupling limit for the heterotic string and $ii)$ because the branes can be sent to infinity in $\mathbb{C}$. In the compact model, the NS5-branes on $\cE$ are sent to $\CZ_+$, signaling the presence of a field theory sector after the blow-up that is, however, decoupled from the local physics at $\CZ_-$. 

Thus, in the heterotic dual of the transition discussed in~\ref{sec:mojodojo}, all three ingredients participating in the F-theory blow-up in a compact model can be interpreted as some form of wrapped, spacetime-filling NS5-branes: The blow-up itself is associated with an NS5-branes wrapping the base curve $C_b\subset X_H$, whereas the change in the number of D3-branes and complex structure deformations is related to the physics of (coincident) NS5-brane wrapping $\cE\subset X_H$. The F-theory transition can hence be viewed in the dual heterotic theory as nucleating and annihilating spacetime-filling NS5-brane. This is reminiscent of the processes proposed in~\cite{Anderson:2022bpo}, though with the crucial difference that on the heterotic side, there are no extremal geometric transitions involved. In this sense, the process is more similar to the smooth nucleation of NS5-branes in $\cN=2$ compactifications of the heterotic string proposed in~\cite{Monnee:2025msf}.

The heterotic dual of the transition discussed in section~\ref{sec:flux} has a simpler heterotic interpretation. There, we only replace NS5-branes wrapping the elliptic fiber of $X_H$ (i.e., spacetime-filling D3-branes in F-theory) with a gauge bundle (i.e., flux in F-theory). Therefore, in the heterotic dual picture, the two transitions discussed in sections~\ref{sec:mojodojo} and~\ref{sec:flux} are thus two sides of the same coin. The flux transition corresponds to the nucleation of heterotic NS5-branes wrapping the elliptic fiber of $X_H$, whereas the transition in section~\ref{sec:mojodojo} creates a heterotic NS5-brane on a curve in the base $B_2$ of $X_H$. Let us stress that we did not work out all the details of the transition in the heterotic dual picture, but only outlined a qualitative picture. For example, even though the heterotic theory gives a unifying picture of the transitions discussed in sections~\ref{sec:mojodojo}, the origin of the enhanced supersymmetry is less clear in these setups. It would be interesting to gain a better understanding of the appearance of locally enhanced supersymmetry in the heterotic theory.

\section{Beyond Supersymmetry Enhancement}\label{ssec:D3moduli}
So far, we focused on transitions involving loci featuring enhanced supersymmetry in the K\"ahler and the complex structure moduli space of $X_4$. In the former case, the supersymmetry enhancement is associated with flop curves not intersecting the anti-canonical divisor of $ B_3$. In contrast, in the complex structure moduli space, the loci featuring enhanced supersymmetry are associated with a supersymmetric flux vacuum, for which the flux superpotential vanishes. In both cases, we saw that, due to the enhanced supersymmetry, there exist additional branches of the moduli space that augment the dimension of the moduli space -- at least in the local models. 

In this section, we consider a specific example of transitions in 4d $\cN=1$ theories of gravity that do not necessarily involve enhanced supersymmetry. Specifically, we focus on the sector of the 4d $\mathcal{N}=1$ chiral moduli space corresponding to the position moduli $\mathbf{x}^{(\alpha)}$, $\alpha = 1, \dots, n_{\mathrm{D3/M2}}$, of spacetime-filling D3/M2-branes. We are particularly interested in transitions that \emph{locally} involve these brane position moduli, in contrast to the transition discussed in section~\ref{sec:mojodojo}, whose local realization did not require presence of spacetime-filling branes. Because D3/M2-branes are present here, the transitions under consideration need not be accompanied by enhanced supersymmetry. Nevertheless, examining them provides a useful comparison with the previously discussed cases. In the following, we revisit the transitions studied in~\cite{Sethi:1996es,Intriligator:2012ue} from a geometric perspective, emphasizing the role of D3/M2-brane moduli.

We again consider transitions in Minkowski space where 4d $\cN=1$ supersymmetry is unbroken, thus requiring 
\begin{equation}
    W=\partial W=0\,. 
\end{equation}
Apart from the contributions of $G_4$-fluxes to the superpotential, there are non-perturbative contributions to the superpotential due to $\frac12$-BPS D3-brane instantons (or M5-brane instantons in the dual M-theory picture). The non-perturbative superpotential takes the schematic form 
\begin{equation}
    W_{\rm n.p.} = \sum_{\mathbf{k}} \cA_{\mathbf{k}}(\psi_i, \mathbf{x}^{(\alpha)}) \exp(-2\pi k^a T_a)\,,
\end{equation}
where the one-loop Pfaffian $\cA$ depends on the complex structure moduli $\psi_i$ and the D3/M2-brane position moduli $\mathbf{x}^{(\alpha)}$.\footnote{In the following, the effects of two/three-form axions to the superpotential that were discussed in section~\ref{ssec:additionaldeformation} do not play any role which is why we suppress them in the field-dependence of $\cA$.} Moreover, the $T_a$ are the complexified K\"ahler moduli of $B_3$ and the vector $\mathbf{k}$ scans over all BPS D3/M5-brane instanton charges. For a Euclidean D3-brane wrapping a divisor $D$ (or Euclidean M5-brane wrapping $\pi^*(D)$) to contribute to the superpotential, it has to have the right amount of fermionic zero modes since otherwise the Pfaffian $\cA$ vanishes identically. The number of zero modes is determined by the geometry of $\pi^*(D)$~\cite{Witten:1996bn} and divisors with the right of zero modes to contribute to the superpotential are known as \textit{rigid}.\footnote{The condition given in~\cite{Witten:1996bn} on the geometry of $\pi^*(D)$ to give two zero modes is in fact neither necessary nor sufficient due to the effects of fluxes and spacetime-filling D3/M2-branes on the zero mode spectrum~\cite{Kallosh:2005gs,Billo:2008sp, Grimm:2011dj,Bianchi:2011qh,Bianchi:2012pn,Bianchi:2012kt}}

For simplicity, let us consider an elliptically fibered Calabi--Yau fourfold $\widehat{X}_4$ such that there is exactly one rigid divisor contributing to the superpotential. This divisor can, for example, be the pullback $\pi^*(E)$ of an exceptional divisor $E\in H_4(\hat B_3)$ obtained by blowing up a generic point $p$ in the base $B_3$ of another fourfold $X_4$. In the following, we assume this to be the case, implying that $E$ has the topology of $\mathbb{P}^2$. The superpotential then has the simple form 
\begin{equation}
    W_{\rm n.p.} = \cA_{\pi^*(E)}\left(\psi_i, \mathbf{x}^{(\alpha)}\right) \exp(-2\pi T_{\pi^*(E)})\,. 
\end{equation}
In the following, we are only interested in the dependence of $\cA_{\pi^*(E)}$ on the D3/M2-brane moduli $\mathbf{x}^{(\alpha)}$. Let us stress that, unlike in section~\ref{ssec:additionaldeformation}, we are blowing up a \emph{point} as opposed to a curve $C_0$ satisfying $C_0\cdot \bar{K}_{B_3}=0$. In the latter case, the divisor $E$ has the topology of $\mathbb{F}_0$ whereas in the case considered here, the topology of $E$ is $\mathbb{P}^2$. Furthermore, as opposed to the case in section~\ref{ssec:additionaldeformation}, the vertical divisor $\pi^*(E)$ does not contain any three-cycles such that there are no zeros of $\cA_{\pi^*(E)}$ induced by specific vevs for the axions associated with these three-cycles. 

The Euler characteristic of $\widehat{X}_4$ is non-vanishing such that we can introduce spacetime-filling M2-branes to cancel the tadpole induced by $\chi(\hat X_4)$. Consider one of these M2-branes for which we denote the position along $\hat X_4$ by $\mathbf{x}^{(1)}\in \mathbb{R}^8$. If we place the M2-brane on the exceptional divisor $\pi^*(E)$ inside $\hat X_4$, the M5-brane instanton on $\pi^*(E)$ has additional fermionic zero modes coming from the M2/M5-intersection~\cite{Ganor:1996pe}. Thus if $\mathbf{x}^{(1)}\in \pi^*(E)$, the Pfaffian $\cA_{\pi^*(E)}$ has to vanish, i.e., 
\begin{equation}
    \cA_{\pi^*(E)}(\mathbf{x}^{(1)})=0\,,\quad \text{if}\quad {\mathbf{x}^{(1)}\in \pi^*(E)}\,. 
\end{equation}
For this configuration to preserve the original $\cN=1$ supersymmetry, the Pfaffian  must also satisfy  
\begin{equation}
    \partial_{\mathbf{x}^{(1)}}   \cA_{\pi^*(E)}(\mathbf{x}^{(1)})=0\,,\quad \text{if}\quad {\mathbf{x}^{(1)}\in \pi^*(E)}\,. 
\end{equation}
In the following, we assume that this is indeed the case. We thus have a supersymmetric configuration for which $W_{\rm n.p.}=0$.

We now argue that there exists another theory with an equidimensional moduli space that is related to the original theory via an extremal 4d $\cN=1$ transition. We therefore notice that for generic M2-brane position moduli, the divisor $\pi^*(E)$ cannot shrink to a point since this regime is dynamically obstructed through the non-zero superpotential that becomes of $\mathcal{O}(1)$ in string units for small $E$. Instead, if $\mathbf{x}^{(1)}\in \pi^*(E)$, we \emph{can} shrink $E$ to a point. Shrinking $E$ can be viewed as the inverse of the transition considered in \cite{Sethi:1996es}. We therefore notice that after shrinking $E$ to a point we obtain a new Calabi--Yau fourfold $X_4$ over a different base $B_3$ which now does not contain any rigid divisors and hence has $W_{\rm n.p.}\equiv 0$. Conversely, the theory on $\hat{X}_4$ can be obtained from that on $X_4$ by blowing up a point $p$ in the base $B_3$. 

As explained in~\cite{Sethi:1996es} the transition from $X_4$ to $\hat X_4$ changes the Euler characteristic. For a smooth elliptic fibration, the Euler characteristic can be expressed as in~\eqref{eq:chifrombase}. Blowing up a point $p\in B_3$ yields 
\begin{equation} 
\int_{B_3} c_1(B_3)^3 - \int_{\hat B_3} c_1(\hat B_3)^3 = 8\,.
\end{equation} 
Since $c_2(\widehat{B}_3)=p^*c_2(B_3)$ the change in the Euler characteristic is simply given by
\begin{equation}
    \frac{\delta \chi}{24} = \frac{\chi(\widehat{X}_4) - \chi(X_4)}{24} = -120\,. 
\end{equation} 
This means that the complex structure moduli space of $X_4$ is augmented compared to $\hat X_4$ since 
\begin{equation}\label{deltahchi}
    \delta h^{3,1} + \delta h^{1,1} - \delta h^{2,1} = \frac{\delta \chi}{6} = -480 \,,
\end{equation}
from which we deduce 
\begin{equation}\label{deltah31}
    \delta h^{3,1}= h^{3,1}(\widehat{X}_4) - h^{3,1}(X_4) \geq -480 -1\,,
\end{equation}
where we used $\delta h^{1,1}=-1$ since we blow down the divisor $E$ to go from $\hat X_4\to {X}_4$. Notice that, unlike for the transition discussed in section~\ref{ssec:additionaldeformation}, here $\delta h^{2,1}=0$ reflecting the fact that we do not require additional axions, as there is no subsector with enhanced supersymmetry in the K\"ahler sector. 

In analogy to our discussion in section~\ref{ssec:additionaldeformation}, we can consider the blow-up of a point $p\in B_3$ also in a local model. In this case, we do not have to worry about the M2-brane tadpole cancellation; yet it is still instructive to understand the physical reason for $\delta h^{2,1}=0$ in this case. From \eqref{deltahchi}, we infer that the change in the number of two-cycles caused by the blow-up induces an M2-brane charge
\begin{equation}\label{deltaqM2}
    \delta h^{1,1} = 1 \quad \Rightarrow \quad \delta q_{M2} = -\frac14\,. 
\end{equation}We argued above that, for the blow-up to be dynamically possible, an M2-brane has to be placed on $\pi^*(E)$. Thus, unlike in the transition discussed in section~\ref{ssec:additionaldeformation}, the transition in this case \emph{has} to involve a spacetime-filling M2-brane. The M2-brane being stuck to the divisor $\pi^*(E)$ has only six instead of the usual eight scalar position moduli. This can be viewed as changing the M2-brane charge by $\delta q_{M2}=-1/4$ consistent with \eqref{deltaqM2}. Thus, compared to blowing up the curve $C_0$ in section~\ref{ssec:additionaldeformation}, the differences of blowing up a point $p\in B_3$ can be summarized as
\begin{enumerate}
    \item The point $p\in B_3$ does not correspond to a subsector with enhanced supersymmetry, such that we do not expect any additional axions present in the full theory, implying $\delta h^{2,1} = 0$. 
    \item The transition $B_3\to \widehat{B}_3$ by blowing up the point $p\in B_3$ involves a spacetime-filling M2-brane required to cancel the non-perturbative contributions to the superpotential. After the transition, the moduli space of the M2-brane has two real dimensions less, reflecting that through the blow-up, the local M2-brane charge gets reduced by $-1/4$. 
\end{enumerate}
The spacetime-filling M2-brane in the genuine $\cN=1$ transition takes over the role of the axions associated to $\delta h^{2,1}=1$ in the transition featuring enhanced supersymmetry. In particular, they locally cancel tadpole charge induced by $\delta h^{1,1}$ and ensure the vanishing of the non-perturbative superpotential as required for the transition to be possible at zero energy and thus to be supersymmetric. 

Returning to the compact model, we infer from \eqref{deltah31} that following the transition from $X_4 \to \hat X_4$ we lose at least $481$ complex structure moduli such that the fourfold $\hat X_4$ is realized at a special locus $\cM_{\rm c.s.}(\hat X_4)$ inside the complex structure moduli space of $X_4$. This is reminiscent of the situation described in section~\ref{ssec:global}. Again, we can use the spacetime-filling M2-branes to accommodate a transition for any value of the complex structure moduli. However, in this case, the dimension of the M2-brane moduli space does not agree with the codimension of $\cM_{\rm c.s.}(\hat X_4)\subset \cM_{\rm c.s.}(X_4)$. 

Instead of spacetime-filling M2-branes, the tadpole can also be canceled through $G_4$-flux. Notice that the locus  $\cM_{\rm c.s.}(\hat X_4)\subset \cM_{\rm c.s.}(X_4)$ can for example correspond to the locus along which multiple four-cycles shrink to zero size, as is the case, e.g., at symmetric points such as Fermat points in a hypersurface Calabi--Yau. Through the transition, we hence replace several four-cycles with volumes controlled by complex structure moduli of $X_4$ with a divisor on $\widehat{X}_4$. Consider the $G_4$-flux dual to the shrinking four-cycles. For the transition between the theory on $X_4$ and $\widehat{X}_4$, the $G_4$-flux has to satisfy 
\begin{equation}
    \frac12 \int_{ X_4} G_4\wedge G_4 = \frac{\delta\chi}{24} =120\,. 
\end{equation}
Since the flux is dual to the four-cycles vanishing at the special locus $\cM_{\rm c.s.}(\hat X_4)$ it has to stabilize $\mathcal{O}(480)$ complex structure moduli. Let us denote the residual c.s. moduli space in the presence of this $G_4$ flux by $\cM_{\rm G_4}$. The relation between flux-induced tadpole and stabilized complex structure moduli is thus 
\begin{equation}
    N_{\rm flux} \simeq \frac{1}{4} h^{3,1}_{\rm stab}\,,
\end{equation}
which is the order of magnitude compatible with the tadpole conjecture~\cite{Bena:2020xrh}. 

The above transition can be summarized as follows. We start from the theory on $\hat X_4$ for which we placed one of the M2-branes on top of $\pi^*(E)$. By blowing down $E$, we then obtain a the theory on the manifold ${X}_4$ with a specific choice for the $G_4$-flux described above. Let us denote the original theory by $\mathbb{T}_E$ and the theory on ${X}_4$ with $G_4$-flux by $\mathbb{T}_{G_4}$. In fact, both theories have a moduli space of the same dimension. To see that, we notice that compared to $\mathbb{T}_{G_4}$ the theory $\mathbb{T}_E$ has one additional K\"ahler modulus 
\begin{equation}
    T_E = \frac12 \int_{\pi^*(E)} J_{\hat X_4}^3 + i \int_{\pi^*(E)} C_6\,,
\end{equation}
where $C_6$ denotes the magnetic dual of the M-theory 3-form $C_3$. Two of the eight real position moduli of the spacetime-filling M2-brane that are not fixed by the flux choice are instead constrained by the condition $\mathbf{x}^{(1)} \in \pi^*(E)$. In contrast, in the theory $\mathbb{T}_{G_4}$, the position of this M2-brane remains unfixed, as it is not required to cancel the contribution of an M5-brane instanton to the superpotential. To conclude, the net change in moduli in transitioning from $\mathbb{T}_{E} \to \mathbb{T}_{G_4}$ is zero.\footnote{This is to be contrasted to the transitions involving subsectors with enhanced supersymmetry for which the moduli space is effectively enhanced when taking into account the non-perturbative transitions.} Thus, the transition discussed here should be viewed as the genuine 4d $\cN=1$ analogue of an extremal transition in 4d $\cN=2$ theories, where one type of modulus is replaced by another. However, unlike for extremal conifold transitions in non-compact Calabi--Yau threefolds, the 4d $\cN=1$ extremal transition does not involve complex structure moduli since we simply replace a K\"ahler modulus by an M2-brane position modulus. 

Notice that by a similar transition as in section~\ref{sec:flux}, the $G_4$-flux in the theory $\mathbb{T}_{G_4}$ can be replaced by spacetime-filling D3-branes. Since the flux-induced superpotential along the locus $\mathcal{M}_{G_4}\subset \cM_{\rm c.s.}(\widehat{X}_4)$ vanishes, the domain wall realizing the transition $\mathbb{T}_{G_4}\to \mathbb{T}_0$ has vanishing tension. As mentioned in~\cite{Intriligator:2012ue}, $\cM_{G_4}$ corresponds to a locus in moduli space along which four-cycles vanish, thus indicating a strongly coupled field theory sector emerging. The nucleation of the spacetime-filling D3-branes is thus a process that must be described within the strongly interacting field theory, analogous to the global version of the transition discussed in section~\ref{ssec:global}.   

\section{Discussion}\label{sec:discussion}
In this work, we have made progress in understanding the non-perturbative nature of four-dimensional $\mathcal {N} = 1$ theories that arise from string theory. Within string theory, non-perturbative effects oftentimes play a crucial role in ensuring consistency of the resulting low-energy effective theories. In theories with extended supersymmetry, the required non-perturbative effects are well understood and ensure a consistent web of string dualities (including strong-weak coupling dualities) or allow for a consistent field theory description across geometric transitions~\cite{Strominger:1995cz,Ooguri:1996me}. The situation in theories with minimal supersymmetry in four dimensions is significantly different due to the lack of reliable techniques for treating string theory non-perturbatively with fewer than eight supercharges. 

To get a handle on non-perturbative effects in 4d $\cN=1$ theories of string theory, we primarily focused on sectors that exhibit local supersymmetry enhancement. The main advantage of this restriction is that the physics associated with enhanced supersymmetry can serve as a guiding principle to infer the existence of novel non-perturbative effects in string theory. For example, locally enhanced supersymmetry requires the light states associated with the subsector with enhanced supersymmetry to fill out complete $\cN=2$ multiplets even though perturbative string theory only provides the fields in $\cN=1$ multiplets. The missing states must then be of non-perturbative origin. 

We applied this line of reasoning to F-theory compactifications on elliptically fibered Calabi--Yau fourfolds. In these setups, subsectors with enhanced supersymmetry are, e.g., associated with shrinkable curves $C_0$ that do not intersect the anti-canonical class of the base, $B_3$, of the fourfold. In the Type IIB orientifold limit of F-theory, these curves hence do not intersect the orientifold locus, and the local geometry resembles a Calabi--Yau threefold. Even though the local geometry is rather simple, string theory requires interesting non-perturbative effects associated with these subsectors to ensure consistency of the resulting low-energy physics. The locally enhanced supersymmetry predicts that the small volume regime for $C_0$ corresponds to the vicinity of the origin of an $\cN=2$ Higgs branch. The light states are hence expected to form a massless $\cN=2$ hypermultiplet and a massive $\cN=2$ vector multiplet that becomes massless at the origin. However, perturbative Type IIB string theory only provides a massless $\cN=1$ chiral multiplet and a massive $\cN=1$ vector multiplet. Accordingly, the missing states have to have a non-perturbative origin. Guided by this physical expectation, we identified the missing degrees of freedom in this work which are associated with blow-ups of the shrinking curve and D3-brane states. Our procedure is similar in spirit to the approach in~\cite{Ooguri:1996me} in that the non-perturbative information about string theory is deduced by imposing the properties of the underlying field theory sector. In~\cite{Ooguri:1996me}, this led to the computation of instanton effects, whereas here we used the properties of the spectrum in the vicinity of an $\cN=2$ field theory to infer missing light states. 

An obvious question is how these results can be applied to better understand the small volume regime of curves that \emph{do} intersect the anti-canonical divisor of the base and hence do not have additional protection from locally enhanced supersymmetry. Here, we can distinguish between curves that are contained in the anti-canonical divisor, and hence have negative intersection with $\bar{K}_{B_3}$, and the curves with positive intersection with $\bar{K}_{B_3}$. Examples of the second kind were already encountered in our analysis as given by the curves $C'$ and $C''$ obtained after blowing up the curve $C_0$. As we saw, strings obtained by wrapping a D3-brane on either of these curves have a finite tension throughout the moduli space. Nonetheless, a finite number of excitations of this string can become massless once the curve volume is of order one in 10d Planck units. It would be exciting to investigate this systematically. An example of curves contained in $\bar{K}_{B_3}$ are flip curves as analyzed in~\cite{Denef:2005mm}. As a main difference to the flop curves studied in this work, the birational factorization of a flip transition cannot be done crepantly inside a Calabi--Yau fourfold due to the appearance of terminal singularities~\cite{Denef:2005mm}. We thus expect the physics in the small volume regime for flip curves to be significantly different from the case studied in this work, reflecting the fact that there is no subsector with enhanced supersymmetry associated with flip curves. Studying this in detail is a natural next step. 

In our analysis in section~\ref{sec:mojodojo}, we encountered interesting global effects arising if the field theory subsector associated with enhanced supersymmetry is embedded in a gravitational theory. In particular, we saw that the number of complex structure moduli of the Calabi--Yau fourfold and the number of spacetime-filling D3-branes differ before and after the curve $C_0$ is blown up. In this work, we proposed a physical interpretation for this by viewing both the complex structure deformations and D3-branes as a strongly coupled field theory sector that arises if the Calabi--Yau fourfold in which we embed the curve $C_0$ is compact. In particular, we interpreted certain degenerations of the complex structure of the Calabi--Yau fourfold as parameters of strongly coupled field theory sectors, and found a correlation with the type of geometric transition that could occur. This was motivated by analogy to F-theory compactifications on elliptic Calabi--Yau threefolds, where certain complex structure degenerations are associated with little string theories in six dimensions. Obtaining a better understanding of the realization of field theory sectors via complex structure degenerations in four dimensions is an exciting avenue for future research. Apart from the change in complex structure deformations, the change in the number of spacetime-filling D3-branes before and after the blow-up highlights the fact that in 4d $\cN=1$ theories arising from F-theory, the geometry of the compactification manifold does not completely determine the low-energy physics, but that the spacetime-filling branes are a key ingredient to characterize the theory entirely. This has to be contrasted to F-theory compactifications on elliptic Calabi--Yau threefolds, where no spacetime-filling branes are needed to characterize the theory. When blowing up the curve $C_0$ in the base of the elliptic Calabi--Yau fourfold in F-theory, we encounter an additional curve providing extra degrees of freedom for the local $\cN=2$ field theory sector but at the same time lose spacetime-filling D3-branes and complex structure deformations. As we argued, in a dual heterotic description of our F-theory setup, the three ingredients taking part in the blow-up have a common origin as spacetime-filling NS5-branes wrapping different curves in the elliptic Calabi--Yau threefold that serves as heterotic compactification manifold. The heterotic version of the blow-up then corresponds to nucleating an NS5-brane on a curve in the base of the threefold while, at the same time, removing NS5-branes on the elliptic fiber. Such a change in the number of spacetime-filling NS5-branes is reminiscent of the transitions studied in~\cite{Anderson:2022bpo} and \cite{Monnee:2025msf}. The main difference of our setup to the one studied in~\cite{Anderson:2022bpo} is that we do not consider a conifold transition in the heterotic Calabi--Yau threefold. Instead, the setups considered in~\cite{Monnee:2025msf} preserve 4d $\cN=2$ supersymmetry though the transitions do not involve geometric transitions on the heterotic side. In both cases, the nucleation of NS5-branes has to be accompanied with a change in the heterotic gauge bundle which we did not discuss in this work. It would be fascinating to gain a deeper understanding of the heterotic dual of the birational factorization in F-theory. In particular, from the heterotic perspective the origin of the enhanced supersymmetry observed in F-theory is not immediately apparent. This is reminiscent of the 6d setup reviewed in section~\ref{sec:6d} in which the supersymmetry enhancement is also not evident from the heterotic perspective and only visible in the dual F-theory formulation of the theory. 

The results of this work highlight that in theories with only minimal supersymmetry in four dimensions, the different sectors of the moduli space cannot be treated independently. In particular, spacetime-filling D3-branes distinguish theories with $\cN=1$ supersymmetry from their counterparts with extended supersymmetry and are a key ingredient to uncover the non-perturbative completion of the low-energy effective theory including additional branches of the moduli space. Whereas the main focus of this work was on subsectors of the 4d $\cN=1$ theory featuring enhanced supersymmetry, we also discussed moduli space transitions that do not rely on such a supersymmetry enhancement. Concretely, we revisited the transitions discussed in~\cite{Sethi:1996es} associated with blow-ups of a point in the base of elliptically fibered Calabi--Yau fourfolds, focusing on the role of spacetime-filling D3-branes. In particular, we discussed similarities and differences to the transitions associated with blow-ups of a curve in the base analyzed in section~\ref{sec:mojodojo}. Most importantly, we demonstrated that for these transitions to be possible, the local blow-up geometry must involve spacetime-filling D3-branes to cancel any contribution to the non-perturbative superpotential that would dynamically obstruct the transition. This has to be contrasted to the transition discussed in section~\ref{sec:mojodojo} where the local version of the transition did not involve any spacetime-filling D3-branes reflecting the enhanced supersymmetry. 

Our analysis of the 4d $\cN=1$ transitions focused on Minkowski setups for which the superpotential and the scalar potential vanish. In particular, no potential was generated for the scalar fields that we considered, which are hence actual moduli of the theory. From a phenomenological perspective, it would be interesting to relax this assumption and allow for non-vanishing superpotentials or scalar potentials. Neither of these cases will feature enhanced supersymmetry, and the latter will break it entirely. Nonetheless, it would be valuable to explore how the non-perturbative insights gained here extend to such setups, a question we hope to revisit in the near future.

\vspace*{.8cm}

       \centerline{\bf \large Acknowledgements}

\vspace*{.5cm}

We thank Lorenzo Paoloni for his collaboration during the early stages of this work. We also thank Severin Lüst and Damian van de Heisteeg for useful discussions.  G.F.C. wishes to acknowledge the hospitality of the II. Institute for Theoretical Physics at the University of Hamburg and the DESY Theory Group
at different stages of this work. G.F.C. is supported through the grants CEX2020-001007-S and PID2021-123017NB-I00, funded by MCIN/AEI/10.13039 /501100011033 and by ERDF A way of making Europe, and by the grant PRE2021-097279 funded by MCIN/AEI/ 10.13039/\\501100011033 and ESF+. MW is supported in part by Deutsche Forschungsgemeinschaft under Germany’s Excellence Strategy EXC 2121 Quantum Universe 390833306, and by Deutsche Forschungsgemeinschaft through the Collaborative Research Center 1624 “Higher Structures, Moduli Spaces and Integrability.”

\bibliography{biblio/papers_Max}
\bibliographystyle{biblio/JHEP}

\end{document}